\documentclass[aps,pre,reprint,twocolumn,superscriptaddress,showpacs]{revtex4-1}
\pdfoutput = 1
\usepackage{graphicx}
\usepackage[caption=false]{subfig}
\usepackage{amsmath}
\usepackage{color}

\usepackage{upgreek}

\newcommand{\mpu}[1]{\textcolor{black}{#1}}

\begin{document}


\title{Forces and torques on rigid inclusions in an elastic environment:\\ resulting matrix-mediated interactions, displacements, and rotations} 

\author{Mate Puljiz}
\email{puljiz@thphy.uni-duesseldorf.de}
\affiliation{Institut f{\"u}r Theoretische Physik II: Weiche Materie, 
Heinrich-Heine-Universit{\"a}t D{\"u}sseldorf, D-40225 D{\"u}sseldorf, Germany}
\author{Andreas M. Menzel}
\email{menzel@thphy.uni-duesseldorf.de}
\affiliation{Institut f{\"u}r Theoretische Physik II: Weiche Materie, 
Heinrich-Heine-Universit{\"a}t D{\"u}sseldorf, D-40225 D{\"u}sseldorf, Germany}

\date{\today}


\begin{abstract}
Embedding rigid inclusions into elastic matrix materials is a procedure of high practical relevance, for instance for the fabrication of elastic composite materials. 
We theoretically analyze the following situation. 
Rigid spherical inclusions are enclosed by a homogeneous elastic medium under stick boundary conditions. Forces and torques are directly imposed from outside onto the inclusions, or are externally induced between them.
The inclusions respond to these forces and torques by translations and rotations against the surrounding elastic matrix. 
This leads to elastic matrix deformations, and in turn results in mutual long-ranged matrix-mediated interactions between the inclusions.
Adapting a well-known approach from low-Reynolds-number hydrodynamics, we explicitly calculate the \textit{displacements and rotations} of the inclusions from the externally imposed or induced \textit{forces and torques}. Analytical expressions are presented as a function of the inclusion configuration in terms of \textit{displaceability and rotateability} matrices. The role of the elastic environment is implicitly included in these relations. That is, the resulting expressions allow a calculation of the induced displacements and rotations directly from the inclusion configuration, without having to explicitly determine the deformations of the elastic environment. 
In contrast to the hydrodynamic case, compressibility of the surrounding medium is readily taken into account.
We present the complete derivation based on the underlying equations of linear elasticity theory. 
In the future, the method will, for example,
be helpful to characterize the behavior of externally tunable elastic composite materials, to accelerate numerical approaches, as well as
to improve the quantitative interpretation of microrheological results.
\end{abstract}

\maketitle

\section{Introduction}\label{Section_1_introduction}

It is safe to say that elastic composite materials are of huge technological importance. This statement is backed by the fact that concrete, the most abundant man-made material on earth \cite{kurtis2015innovations}, is frequently composed of a cement matrix supported by more rigid particulate inclusions \cite{li1999effective,constantinides2004effect,landis2009explicit,kurtis2015innovations,mechtcherine2014simulation}. Understanding the mutual interactions between the inclusions as well as between the inclusions and the matrix is crucial to understand the overall material performance.

While hardened concrete is a relatively rigid substance, polymeric gel matrices or biological tissue can provide softer elastic environments. 
Then, larger-scale displacements and rotations of embedded inclusions can be observed when  forces and/or torques are externally imposed or induced. Magnetic microrheology observes the displacements of probe particles caused by externally applied magnetic field gradients \cite{ziemann1994local,bausch1999measurement,waigh2005microrheology,wilhelm2008out}. 
For instance, the mechanical response of the cytoskeleton \cite{ziemann1994local,bausch1999measurement,waigh2005microrheology,wilhelm2008out,mizuno2007nonequilibrium} was analyzed in this way. Similarly, the rotational motion of magnetic rods under externally imposed magnetic torques can be used for microrheological purposes \cite{bender2011synthesis,roeder2012shear,bender2013determination}. The same is true for tracking the relative displacements between particles that respond to mutual magnetic forces induced between them \cite{puljiz2016forces}. 

Thinking of rigid inclusions embedded in a soft elastic polymeric gel matrix, artificial soft actuators represent a natural type of application \cite{an2003actuating,fuhrer2009crosslinking,bose2012soft,roussel2014electromechanical}. Different approaches are possible. On the one hand, a net external force or torque can be imposed onto the inclusions. For example, magnetic particles are drawn towards external field gradients \cite{zrinyi1996deformation}, while anisotropic particles may experience a torque under an external electric or magnetic field \cite{stolbov2011modelling,ilg2013stimuli,weeber2015ferrogels,cremer2016superelastic}. In these cases, the externally imposed forces or torques are transmitted by the inclusions to the embedding matrix and lead to overall deformations. 
On the other hand, genuinely electrostrictive or magnetostrictive effects can be exploited when external electric or magnetic fields induce mutual attractions and repulsions between the embedded inclusions and in total lead to macroscopic deformations \cite{diguet2009dipolar,wongtimnoi2011improvement,allahyarov2015simulation}. In addition to that, the overall mechanical properties can be tuned from outside by external fields in such materials. This allows, during application, to reversibly adjust from outside the elastic properties to a current need. Examples are the magnitudes of the elastic moduli \cite{jolly1996magnetoviscoelastic,an2003actuating,gao2004electrorheological,minagawa2005electro,filipcsei2007magnetic,ivaneyko2012effects,menzel2015tuned,odenbach2016microstructure}, nonlinear stress-strain behavior \cite{cremer2015tailoring,cremer2016superelastic}, or dynamic properties \cite{jarkova2003hydrodynamics,bohlius2004macroscopic,tarama2014tunable,pessot2016dynamic,volkova2016motion}, allowing for instance for the construction of tunable soft damping devices \cite{deng2006development,sun2008study,li2008research}.

In all these situations, for a theoretical characterization and quantitative description of the material behavior, it is necessary to determine the induced displacements and rotations of the rigid inclusions. This is a 
many-body problem. The inclusions are enclosed by the elastic matrix and transmit the forces and torques to their embedding environment. As a consequence, the matrix gets deformed. The other inclusions are exposed to these induced deformations of their environment. As a consequence, they are additionally displaced and rotated. Moreover, the inclusions are rigid and resist deformations that would result from the induced matrix deformations. This resistance leads to further stresses on the embedding matrix and in turn to additional \textit{matrix-mediated interactions} between the inclusions. 

One can address this problem using simplified representations of the surrounding matrix, e.g., in elastic-spring \cite{pessot2014structural} or elastic-rod \cite{biller2014modeling,biller2015mesoscopic} models. Alternatively, one can directly perform complete finite-element simulations \cite{han2013field,spieler2013xfem,attaran2016modeling,metsch2016numerical} or apply related schemes of simulation \cite{cremer2015tailoring,cremer2016superelastic} to explicitly cover the matrix behavior.

Here, for rigid spherical particles embedded with stick boundary conditions in the elastic matrix, we explicitly solve the problem analytically. Following the above cause-and-effect principle, we start from the forces and torques acting on the embedded particles. We then calculate the resulting coupled displacements and rotations of all particles, including the described matrix-mediated interactions between them. Our analytical results are given in terms of \textit{displaceability and rotateability matrices} that, when multiplied with the forces and torques, lead to the caused displacements and rotations. These expressions solely depend on the configuration of the inclusions and \textit{implicitly contain the role of the elastic environment}. As a strong benefit, the deformations of the elastic environment do not need to be calculated explicitly any more. Therefore, in the future, one can directly calculate analytically the resulting displacements and rotations of the inclusions, without needing to resolve the induced elastic matrix deformations any longer. (To avoid confusion, we note that the term ``matrix'' is used both for the elastic environment as well as for the mathematical representation of second-rank tensors).

Our approach is based on the fact that for the static linear elasticity equations a Green's function is available \cite{landau1986theory}.
We then adapt a method from low-Reynolds-number hydrodynamics, called the \textit{method of reflections} \cite{karrila1991microhydrodynamics,dhont1996introduction}. There, hydrodynamic interactions, i.e., fluid flows induced by suspended particles, play the role of the matrix-mediated interactions in our case. In hydrodynamics, the approach turned out to be extremely successful in characterizing the behavior of suspensions of colloidal particles \cite{felderhof1977hydrodynamic,ermak1978brownian,durlofsky1987dynamic,zahn1997hydrodynamic,meiners1999direct,dhont2004thermodiffusion,rex2008influence,jager2013dynamics}, i.e., nano- to micrometer-sized objects, and of self-propelled microswimmers \cite{pooley2007hydrodynamic,lauga2009hydrodynamics,baskaran2009statistical,menzel2016dynamical}. We expect similar benefits for the characterization of elastic composite materials in the future. In contrast to the hydrodynamic case, \textit{compressible} elastic matrices are readily described as well.

Technically, the method corresponds to an iterative procedure in orders of the inverse separation distance between the rigid inclusions. We here proceed to the fourth order in this inverse distance, but in principle one can proceed to arbitrary order. Parts of our results were presented before (for instance, the elastic Fax\'{e}n laws \cite{kim1995faxen,norris2008faxen}, see below, the derivation of which we here, however, present by explicit calculation in analogy to the hydrodynamic procedure in Refs.~\onlinecite{dhont1996introduction} and \onlinecite{batchelor1972hydrodynamic}). Mostly, in the very few previous approaches on this subject, the displacements were used as a starting point, and expressions for the forces and torques necessary to achieve these displacements were then derived \cite{phanthien1994loadtransfer,kim1995faxen}. Here, we follow the converse route, i.e., the forces and torques are used as known input, and we then calculate the resulting displacements and rotations. 
This is in agreement with the cause-and-effect chain that usually applies in experiments.
Our presentation has two main purposes. First, we provide more explicitly the steps of derivation outlined already in Ref.~\onlinecite{puljiz2016forces} for the displaceability matrix. Second, we amend this procedure by the rotational component, so that now also the influence of imposed torques and the couplings between translational and rotational degrees of freedom are included.

We start in Sec.~\ref{Section_2_greens_function} with a brief overview on the underlying equations of linear elasticity theory, including the corresponding Green's solution.
In Sec.~\ref{Section_3_multipole_expansion}, we review the multipole expansion (a Taylor expansion) of the Green's solution around the center of a rigid inclusion.
Subsequently, the calculation of the displacement field around a finite-sized sphere subject to an external force or torque is explicitly described in Secs.~\ref{Section_4_uniformly_translated_sphere} and \ref{Section_5_uniformly_rotated_sphere}, respectively.
In Sec.~\ref{Section_6_faxen_laws}, the derivation of the translational and rotational Fax\'en laws of elasticity is presented explicitly; these expressions describe how a single spherical inclusion is displaced and rotated in a given, imposed matrix deformation. The Fax\'en laws enable us in Secs.~\ref{Section_7_displaceability_rotateability}--\ref{Section_9_three_spheres} to calculate the mutual matrix-mediated interactions between spherical inclusions in elastic media.
They contribute to the \textit{displaceability and rotateability matrices} defined in Sec.~\ref{Section_7_displaceability_rotateability}, which allow to directly calculate from given \textit{forces and torques} on all inclusions their coupled \textit{displacements and rotations}.
We explicitly calculate the components of these matrices to fourth order in inverse inclusion separation distance. For this purpose, we first restrict ourselves to two-sphere interactions in Sec.~\ref{Section_8_two_rigid_sphere_interaction} and after that include three-sphere interactions in Sec.~\ref{Section_9_three_spheres}. 
Parts of our results are briefly illustrated by considering simplified and idealized example situations in Sec.~\ref{Section_10_rotations}.
Brief conclusions and a short outlook follow in Sec.~\ref{Section_11_conclusions}, while several technical details are added in the Appendices to render the presentation fully self-contained.

\section{Green's function in linear elasticity theory}\label{Section_2_greens_function}

Throughout, we consider an isotropic, homogeneous, and infinitely extended elastic matrix. 
Displacements of the volume elements of the elastic matrix are described by the displacement field $\mathbf{u}(\mathbf{r})$. 
We consider a point force $\mathbf{F}$ acting on the matrix at position $\mathbf{r}_0$. 
If the deformations are restricted to the linear regime, then 
$\mathbf{u}(\mathbf{r})$ obeys the Navier-Cauchy equations \cite{navier1822memoire} of linear elasticity theory,
\begin{equation}\label{navier_cauchy}
	\nabla^2\mathbf{u}(\mathbf{r}) + \frac{1}{1-2\nu}\nabla\nabla\cdot\mathbf{u}(\mathbf{r}) ={} -\frac{1}{\mu}\mathbf{F}\delta(\mathbf{r}-\mathbf{r}_0),
\end{equation}
with $\nu$ the Poisson ratio connected to the matrix compressibility, $\mu$ the shear modulus, and $\delta(\mathbf{r})$ the Dirac delta function.

At positions different from $\mathbf{r}_0$, three relations arise from Eq.~(\ref{navier_cauchy}) that will prove to be useful in subsequent sections.
First, taking the divergence of Eq.~(\ref{navier_cauchy}), we obtain (for $\mathbf{r}\neq\mathbf{r}_0$)
\begin{equation}
	\nabla^2\nabla\cdot\mathbf{u} ={}0.
\end{equation}
Second, working on Eq.~(\ref{navier_cauchy}) with $\nabla^2$ therefore leads to
\begin{equation}\label{biharmonic}
	\nabla^4\mathbf{u}={} \mathbf{0},
\end{equation}
which is referred to as biharmonic equation.
The third relation is obtained by taking the 
curl
of Eq.~(\ref{navier_cauchy}), resulting in
\begin{equation}\label{rot_u}
	\nabla\times\nabla^2\mathbf{u}={}\mathbf{0}.
\end{equation}

The general solution of Eq.~(\ref{navier_cauchy}) can be expressed by a Green's function,
\begin{equation}\label{u_G_F}
	\mathbf{u}(\mathbf{r}) ={} \mathbf{\hspace{.02cm}\underline{\hspace{-.02cm}G}}(\mathbf{r},\mathbf{r}_0) \cdot \mathbf{F},
\end{equation}
with $\mathbf{\hspace{.02cm}\underline{\hspace{-.02cm}G}}(\mathbf{r},\mathbf{r}_0)$ a tensor of rank 2 (we mark second-rank tensors and matrices by an underscore).
Due to the homogeneity and isotropy of the material, $\mathbf{\hspace{.02cm}\underline{\hspace{-.02cm}G}}(\mathbf{r},\mathbf{r}_0)$ is a function of the vector $\mathbf{r}-\mathbf{r}_0$ only. 
For completeness, we briefly reproduce its derivation (see, e.g., Ref.~\onlinecite{weinberger2005lecture}).

The generalized Hooke's law \cite{landau1986theory} of linear elasticity theory reads 
\begin{equation}\label{stress_stiffness_strain}
	\sigma_{kp} ={} \lambda_{kpim}u_{im},
\end{equation}
with $\sigma_{kp}$ and $u_{im}$ the components of the stress and strain tensor, respectively. $\lambda_{kpim}$ summarizes the elastic coefficients, and the Einstein summation rule is applied.
For isotropic materials, the tensor of elastic coefficients takes the form \cite{landau1986theory}
\begin{equation}\label{clandau}
	\lambda_{kpim} ={} \lambda \delta_{kp}\delta_{im} + \mu (\delta_{ki}\delta_{pm} + \delta_{km}\delta_{pi}),
\end{equation}
with
\begin{equation}
	\lambda ={} \frac{2\mu\nu}{1-2\nu},
\end{equation}
whereas the linearized strain tensor \cite{landau1986theory} reads
\begin{equation}\label{def_linear_strain_tensor}
	u_{im} = \frac{1}{2}\left(\nabla_i u_m + \nabla_m u_i\right).
\end{equation}

\begin{figure}
\centerline{\includegraphics[width=\columnwidth]{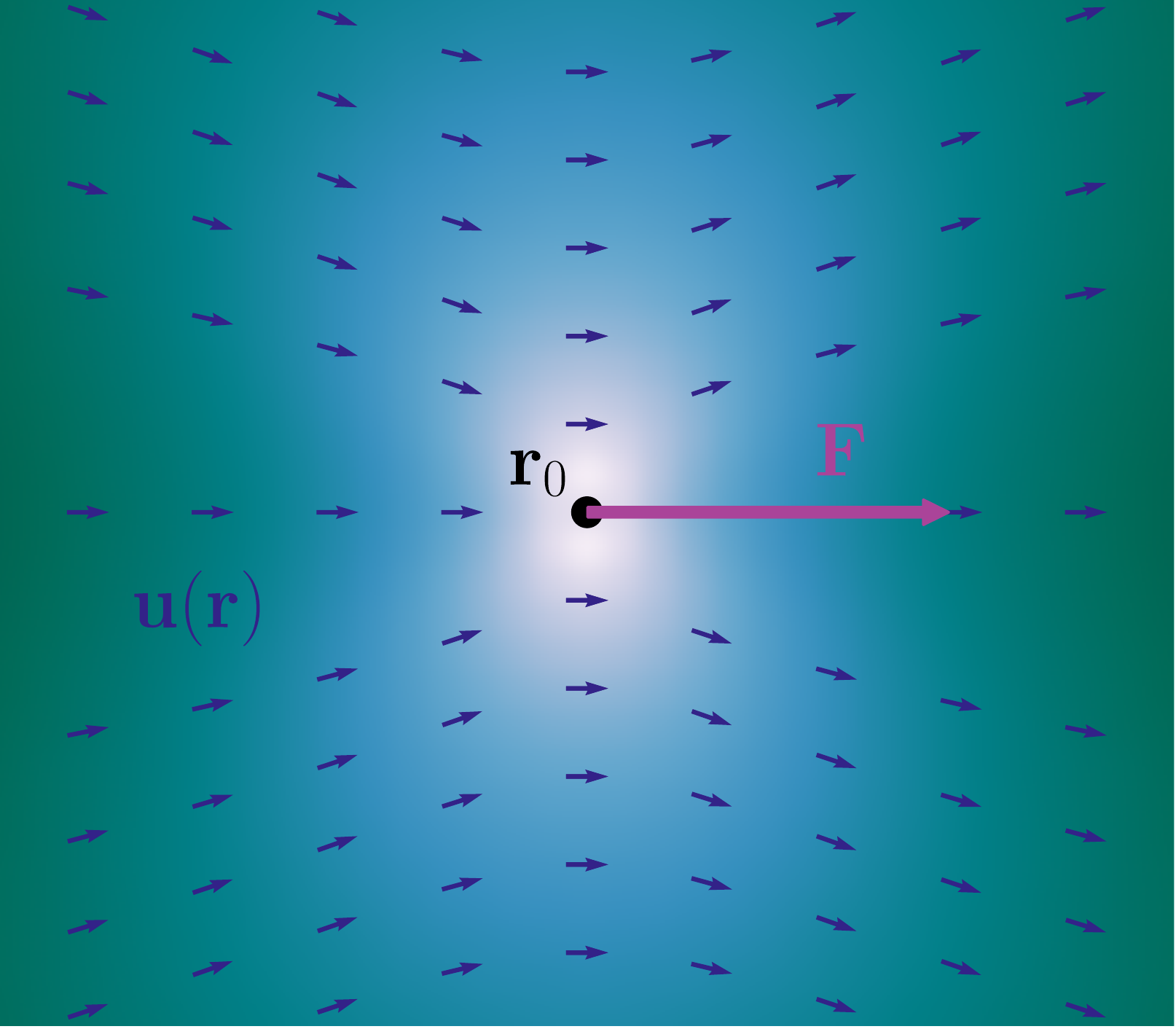}}
\caption{Illustration of the displacement field $\mathbf{u}(\mathbf{r})$ generated by a point force $\mathbf{F}$ acting on the matrix at position $\mathbf{r}_0$. The displacement field is obtained from Eq.~(\ref{u_G_F}) via the elastic Green's function in Eq.~(\ref{greens_function}). Small arrows, for visibility rescaled to identical length, indicate the direction of the displacement field, whereas the background color represents the local magnitude of $\mathbf{u}(\mathbf{r})$ on a logarithmic scale. The brighter the color, the higher the magnitude of $\mathbf{u}(\mathbf{r})$.}
\label{FIG_GREENS_FUNCTION}
\end{figure}

We assume an arbitrary simply connected volume $V$ of the elastic material.
The only force acting on this material is our point force $\mathbf{F}$ at position $\mathbf{r}_0$.
In equilibrium, this point force is balanced by the forces resulting from the surface stress:
\begin{eqnarray}
	\int_{\partial V}\mathrm{d}S_p \sigma_{kp} + F_k ={} 0.
\end{eqnarray}
Using the Gaussian divergence theorem, the surface integral can be converted into a volume integral. Therefore, inserting Eqs.~(\ref{u_G_F}), (\ref{stress_stiffness_strain}), and (\ref{def_linear_strain_tensor}) yields the expression
\begin{equation}
	\int_V \mathrm{d}V \Bigg[\lambda_{kpim}\nabla_m\nabla_p G_{ij}(\mathbf{r}-\mathbf{r}_0) + \delta_{jk}\delta(\mathbf{r}-\mathbf{r}_0)\Bigg] F_j ={} 0.
\end{equation}
Since the above equation must hold true for any arbitrary volume and point of attack $\mathbf{r}_0$, the Green's function $G_{ij}(\mathbf{r}-\mathbf{r}_0)$ must satisfy the equilibrium condition
\begin{equation}\label{ggb}
	\lambda_{kpim}\nabla_m\nabla_pG_{ij}(\mathbf{r}-\mathbf{r}_0) + \delta_{jk}\delta(\mathbf{r}-\mathbf{r}_0)={}0.
\end{equation}
This equation can be solved by Fourier forth and back transformation, see Appendix A, resulting in
\begin{equation}\label{greens_function}
	\mathbf{\hspace{.02cm}\underline{\hspace{-.02cm}G}}(\mathbf{r}) ={} \frac{1}{16\pi(1-\nu)\mu}\Bigg[\frac{3-4\nu}{r}\mathbf{\underline{\hat{I}}} + \frac{\mathbf{r}\mathbf{r}}{r^3}\Bigg],
\end{equation}
with $\mathbf{\underline{\hat{I}}}$ the identity matrix and $\mathbf{r}\mathbf{r}$ a dyadic product.
A graphical representation of Eqs.~(\ref{u_G_F}) and (\ref{greens_function}) is given in Fig.~\ref{FIG_GREENS_FUNCTION}.
For incompressible materials (in the regime of linear elasticity), $\nu$ takes the value $1/2$. 
In this case, the Green's function in Eq.~(\ref{greens_function}) has the same form as the Oseen tensor in low-Reynolds-number hydrodynamics \cite{landau1987fluid,karrila1991microhydrodynamics,dhont1996introduction}, where the hydrodynamic viscosity takes the place of $\mu$.
In general, $\mathbf{\hspace{.02cm}\underline{\hspace{-.02cm}G}}(\mathbf{r})$ used in Eq.~(\ref{u_G_F}) solves Eq.~(\ref{navier_cauchy}).

\section{Multipole expansion}\label{Section_3_multipole_expansion}
Using the elastic Green's function $\mathbf{\hspace{.02cm}\underline{\hspace{-.02cm}G}}(\mathbf{r})$, we can express the matrix displacement field $\mathbf{u}(\mathbf{r})$ generated by an arbitrarily shaped embedded rigid particle centered at the origin as
\begin{equation}\label{dis_rep}
	\mathbf{u}(\mathbf{r}) ={} \int_{\partial V}  \mathrm{d}S^\prime \mathbf{\hspace{.02cm}\underline{\hspace{-.02cm}G}}(\mathbf{r}-\mathbf{r}^\prime)\cdot\mathbf{f}(\mathbf{r}^\prime).
\end{equation}
Here  $\mathbf{r}^\prime$ is located on the particle surface $\partial V$ and $\mathbf{f}(\mathbf{r}^\prime)$ is the force per unit area exerted by the rigid particle onto the matrix.
This equation expresses a superposition of displacement fields generated by point forces on the particle surface. A similar situation arises in electrostatics, where a localized continuous charge distribution can be expressed as a superposition of point charges, each of which contributing to the overall electric potential.
Moreover, similarly to the electrostatic potential of point charges, in Eq.~(\ref{greens_function}) we have $\mathbf{\hspace{.02cm}\underline{\hspace{-.02cm}G}}(\mathbf{r})\sim r^{-1}$.
Therefore, it is possible to perform a multipole expansion of the Green's function.
This is well-known for low-Reynolds-number hydrodynamics \cite{karrila1991microhydrodynamics} and has previously been adapted to elastostatics \cite{phanthien1993rigid}. 
We follow the procedure as described for the hydrodynamic case in Ref.~\onlinecite{karrila1991microhydrodynamics}.

In the far field, one has $|\mathbf{r}|\gg|\mathbf{r}^\prime|$ in Eq.~(\ref{dis_rep}). 
The Taylor series of $\mathbf{\hspace{.02cm}\underline{\hspace{-.02cm}G}}(\mathbf{r}-\mathbf{r}^\prime)$ in $\mathbf{r}^\prime$ around $\mathbf{r}^\prime=\boldsymbol{0}$ reads
\begin{equation}\label{taylor}
G_{ij}(\mathbf{r}-\mathbf{r}^\prime)
	={} \sum_{n=0}^{\infty}\frac{(-1)^n}{n!}(\mathbf{r}^\prime\cdot\nabla)^nG_{ij}(\mathbf{r}).\qquad
\end{equation}
Inserting Eq.~(\ref{taylor}) into Eq.~(\ref{dis_rep}), we obtain the components of the displacement field as
\begin{eqnarray}
	u_i(\mathbf{r}) &={} &\sum_{n=0}^{\infty}\frac{(-1)^n}{n!}
	 \int_{\partial V} \mathrm{d}S^\prime f_j(\mathbf{r}^\prime)\,(\mathbf{r}^\prime\cdot\nabla)^n G_{ij}(\mathbf{r})\notag \\
	&={} &G_{ij}(\mathbf{r})F_j - \frac{\partial G_{ij}(\mathbf{r})}{\partial r_k}D_{jk}+...\label{multipole_general}
\end{eqnarray}
with
\begin{eqnarray}
	F_j ={} \int_{\partial V}\mathrm{d}S'f_j(\mathbf{r}^\prime), \quad
	D_{jk} ={} \int_{\partial V}\mathrm{d}S'f_j(\mathbf{r}^\prime)r_{k}^\prime.
\end{eqnarray}
Here, $\mathbf{F}$ can be identified as the total force that the particle exerts on the matrix. The $\mathbf{\underline{D}}$-tensor can be split into an antisymmetric and a symmetric part,
\begin{equation}
	D_{jk} ={} T_{jk} + S_{jk},
\end{equation}
with
\begin{eqnarray}
	T_{jk} &={} &\frac{1}{2}\int_{\partial V}\mathrm{d}S^\prime [f_j(\mathbf{r}^\prime)r_{k}^\prime- f_k(\mathbf{r}^\prime)r_j^\prime],\\
	S_{jk} &={} &\frac{1}{2}\int_{\partial V}\mathrm{d}S^\prime [f_j(\mathbf{r}^\prime)r_{k}^\prime+ f_k(\mathbf{r}^\prime)r_j^\prime].\label{stresslet}
\end{eqnarray}
The symmetric tensor $S_{jk}$ is called stresslet.
Furthermore, we set the components of the torque $\mathbf{T}$ that the particle exerts on the matrix to
\begin{equation}
	T_i :={} \epsilon_{ijk}\int_{\partial V}\mathrm{d}S^\prime r_{j}^\prime f_k(\mathbf{r}^\prime) ={} -\epsilon_{ijk} T_{jk},\label{torque}
\end{equation}
with $\epsilon_{ijk}$ the Levi-Civita symbol.
Therefore, we can express the corresponding part in Eq.~(\ref{multipole_general}) through
\begin{equation}
	T_{jk}\frac{\partial G_{ij}}{\partial r_k}={} -\frac{1}{2}\epsilon_{jkl}T_l \frac{\partial G_{ij}}{\partial r_k}={} \frac{1}{2}(\mathbf{T}\times\nabla)_j G_{ij}.
\end{equation}

In sum, we obtain the following expression for the first terms of the multipole expansion,
\begin{equation}
	\mathbf{u}(\mathbf{r}) ={} \mathbf{\hspace{.02cm}\underline{\hspace{-.02cm}G}}(\mathbf{r})\cdot\mathbf{F} - \left(\frac{1}{2}\mathbf{T}\times\nabla +  \mathbf{\underline{S}}\cdot\nabla\right) \cdot \mathbf{\hspace{.02cm}\underline{\hspace{-.02cm}G}}(\mathbf{r})\label{multipole_fts},
\end{equation}
which corresponds to the displacement field around a rigid particle in far-field approximation.

\section{Displacement field induced by a uniformly translated rigid spherical inclusion}\label{Section_4_uniformly_translated_sphere}
To facilitate our analytical approach, we now confine ourselves to rigid spherical particles embedded in the elastic matrix.
The center of such a sphere of volume $V$ is located at position $\mathbf{r}_0$ and $a$ is its radius.
If an external force $\mathbf{F}$ uniformly translates the sphere, it creates a displacement field in the surrounding matrix.
Assuming that the elastic matrix sticks to the surface $\partial V$ of the sphere and that the displacement field vanishes at infinity,
the boundary conditions for $\mathbf{u}(\mathbf{r})$ follow as
\begin{equation}\label{boundary_sphere}
	\mathbf{u}(\mathbf{r}\in\partial V) ={} \mathbf{U},\qquad
	\mathbf{u}(|\mathbf{r}|\rightarrow\infty) ={} \mathbf{0}.  
\end{equation}
Here $\mathbf{U}$ is the translation of the sphere caused by the external force, which due to the particle rigidity simultaneously applies for all its surface points.

The resulting displacement field can be expressed in terms of the elastic Green's function $\mathbf{\hspace{.02cm}\underline{\hspace{-.02cm}G}}(\mathbf{r}-\mathbf{r}_0)$, 
see Eq.~(\ref{dis_rep}). 
The integral in Eq.~(\ref{dis_rep}), summing over all the contributions from the point forces on the particle surface at positions $\mathbf{r}^\prime\in\partial V$, can for a sphere be calculated explicitly, see Ref.~\onlinecite{dhont1996introduction} for the case of low-Reynolds-number hydrodynamics. However, this is a lengthy calculation, and we follow the elegant approach outlined in Refs.~\onlinecite{karrila1991microhydrodynamics,phanthien1993rigid}.

Due to the linearity of the Navier-Cauchy equations Eq.~(\ref{navier_cauchy}), there is only one unique solution satisfying the prescribed boundary conditions. 
Assuming $\mathbf{F}\sim
\mathbf{U}$ in the linear regime, an ansatz $\mathbf{u}(\mathbf{r})\sim\mathbf{\hspace{.02cm}\underline{\hspace{-.02cm}G}}(\mathbf{r}-\mathbf{r}_0)\cdot\mathbf{F}\sim\mathbf{\hspace{.02cm}\underline{\hspace{-.02cm}G}}(\mathbf{r}-\mathbf{r}_0)\cdot\mathbf{U}$ appears plausible.
Moreover, since on $\partial V$ the displacement field $\mathbf{u}(\mathbf{r})\sim\mathbf{\hspace{.02cm}\underline{\hspace{-.02cm}G}}(\mathbf{r}-\mathbf{r}_0)\cdot\mathbf{U}$ must satisfy Eq.~(\ref{boundary_sphere}), on $\partial V$ the overall multiplicand of $\mathbf{U}$ in this expression must be proportional to $\mathbf{\underline{\hat{I}}}$.
This is accomplished by an additional differential operator acting on $\mathbf{\hspace{.02cm}\underline{\hspace{-.02cm}G}}(\mathbf{r}-\mathbf{r}_0)$,
\begin{equation}
	\left(1+\frac{a^2}{6}\nabla^2\right)\mathbf{\hspace{.02cm}\underline{\hspace{-.02cm}G}}(\mathbf{r}-\mathbf{r}_0)\bigg|_{|\mathbf{r}-\mathbf{r}_0|=a} ={}\frac{5-6\nu}{24\pi(1-\nu)\mu a}\mathbf{\underline{\hat{I}}}.
\end{equation}
Altogether, 
\begin{equation}
	\mathbf{u}(\mathbf{r}) ={} \frac{24\pi(1-\nu)\mu a}{5-6\nu}\left(1+\frac{a^2}{6}\nabla^2\right)\mathbf{\hspace{.02cm}\underline{\hspace{-.02cm}G}}(\mathbf{r}-\mathbf{r}_0)\cdot\mathbf{U}
\label{displacement_sphere_v}
\end{equation}
satisfies the boundary conditions Eq.~(\ref{boundary_sphere}) as well as Eq.~(\ref{navier_cauchy}) and thus, due to the uniqueness of the solution, is the desired result.

For $a\rightarrow 0$ and $|\mathbf{r}-\mathbf{r}_0|>a$, the contribution $\frac{a^2}{6}\nabla^2$ becomes negligible and we must reproduce Eq.~(\ref{u_G_F}).
In this way, we find
\begin{equation}\label{force_translation}
	\mathbf{F} ={} \frac{24\pi(1-\nu)\mu a}{5-6\nu} \mathbf{U}
\end{equation}
or, equivalently,
\begin{equation}
	\mathbf{u}(\mathbf{r}\in\partial V) ={} \mathbf{U} ={} \frac{5-6\nu}{24\pi(1-\nu)\mu a}\mathbf{F}.\label{stokes}
\end{equation}
As a consequence, we may rewrite Eq.~(\ref{displacement_sphere_v}) as
\begin{equation}\label{displacement_sphere}
	\mathbf{u}(\mathbf{r}) ={} \left(1+\frac{a^2}{6}\nabla^2\right)\mathbf{\hspace{.02cm}\underline{\hspace{-.02cm}G}}(\mathbf{r}-\mathbf{r}_0)\cdot\mathbf{F}.
\end{equation}
This is the elastic analogue to the hydrodynamic Stokes flow \cite{dhont1996introduction}.

Since, as we just argued, the solution \mpu{in} Eq.~(\ref{displacement_sphere}) is exact, we can for a spherical particle insert it into Eq.~(\ref{dis_rep}) to find for $|\mathbf{r}-\mathbf{r}_0|\ge a$ the relation
\begin{equation}\label{sphere_relation}
	\int_{\partial V}  \mathbf{\hspace{.02cm}\underline{\hspace{-.02cm}G}}(\mathbf{r}-\mathbf{r}^\prime)\cdot \mathbf{f}(\mathbf{r}^\prime) \mathrm{d}S^\prime ={} \left(1+\frac{a^2}{6}\nabla^2\right)\mathbf{\hspace{.02cm}\underline{\hspace{-.02cm}G}}(\mathbf{r}-\mathbf{r}_0)\cdot\mathbf{F},
\end{equation}
which we will need later.

\section{Displacement field induced by a uniformly rotated rigid spherical inclusion}\label{Section_5_uniformly_rotated_sphere}
In a similar way, we can ask for the displacement field generated in an elastic matrix by a uniformly rotated rigid spherical inclusion at position $\mathbf{r}_0$.
For this purpose, we consider an external torque $\mathbf{T}$ acting on the inclusion (see Refs.~\onlinecite{karrila1991microhydrodynamics,dhont1996introduction} for the low-Reynolds-number hydrodynamic and Ref.~\onlinecite{phanthien1993rigid} for the elastic case). 
The rotation of the particle is quantified by the absolute (static) rotation vector $\boldsymbol{\Omega}$.
Then the boundary conditions on the surface $\partial V$ of the particle and at infinity read
\begin{equation}
	\mathbf{u}(\mathbf{r}\in\partial V) ={}\boldsymbol{\Omega}\times(\mathbf{r}-\mathbf{r}_0), \qquad \mathbf{u}(|\mathbf{r}|\rightarrow\infty) ={} \mathbf{0}.
\end{equation}
Inserting the displacement field
\begin{equation}\label{displacement_rotated_sphere}
	\mathbf{u}(\mathbf{r}) ={} \left(\frac{a}{|\mathbf{r}-\mathbf{r}_0|}\right)^{\!\!3}\boldsymbol{\Omega}\times(\mathbf{r}-\mathbf{r}_0)
\end{equation}
into these boundary conditions as well as into Eq.~(\ref{navier_cauchy}) confirms that it is the unique solution of the problem. As will be shown in Sec.~\ref{Section_6_faxen_laws}, see Eq.~(\ref{faxen_w}), the torque that is externally imposed on the inclusion is related to the rotation vector $\boldsymbol{\Omega}$ via
\begin{equation}\label{torque_rotation}
	\mathbf{T}={}8\pi\mu a^3\boldsymbol{\Omega},
\end{equation}
with $a$ the radius of the sphere.

\section{Fax\'en's laws}\label{Section_6_faxen_laws}

In low-Reynolds-number hydrodynamics, Fax\'en's laws describe how a spherical particle is translated, rotated, and which stresses act onto it in an imposed fluid flow \cite{batchelor1972hydrodynamic,dhont1996introduction,karrila1991microhydrodynamics}. The fluid is typically considered as incompressible.

Due to the similarities of the underlying equations, the procedure can be transferred to the elastic case. 
That is, we now consider an (externally) imposed deformation of our elastic matrix as described by a displacement field $\mathbf{u}(\mathbf{r})$. 
We then calculate how a rigid spherical particle embedded in the elastic matrix and exposed to this displacement field is translated, rotated, and which stresses act onto it. 
A possible \textit{compressibility} of the elastic matrix is readily included. Such elastic Fax\'en laws have been outlined before \cite{kim1995faxen,norris2008faxen}. Here, we present an explicit derivation by direct calculation. We adapt the hydrodynamic approach in Refs.~\onlinecite{dhont1996introduction} and \onlinecite{batchelor1972hydrodynamic} by transferring it to the elastic case.

We consider a rigid spherical inclusion of radius $a$ embedded in the elastic matrix at position $\mathbf{r}_0$. 
In addition to the displacement field imposed onto the matrix, the embedded particle may still be subject to external forces or torques. Moreover, its rigidity resists the imposed matrix deformations. Therefore, its surface elements exert additional forces onto the matrix, summarized again by the surface force density $\mathbf{f}(\mathbf{r}')$ with $\mathbf{r}'\in\partial V$ and $\partial V$ the surface of the particle.
The additional displacement field resulting from $\mathbf{f}(\mathbf{r}')$ is calculated according to Eq.~(\ref{dis_rep}). 
Due to the linearity of Eq.~(\ref{navier_cauchy}), the different contributions to the overall displacement field simply superimpose. 
Describing again translations and rotations of the sphere by a translation vector $\mathbf{U}$ and a (static) rotation vector $\boldsymbol{\Omega}$, respectively, we obtain in total for the surface points $\mathbf{r}\in\partial V$ the stick boundary condition
\begin{equation}
	 U_i + [\boldsymbol{\Omega}\times(\mathbf{r}-\mathbf{r}_0)]_i ={}  \int_{\partial V} G_{ij}(\mathbf{r}-\mathbf{r}^\prime)f_j(\mathbf{r}^\prime) \mathrm{d}S^\prime + u_i(\mathbf{r})\label{faxen_grund}.
\end{equation}

On the left-hand side of this equation, we find the displacements of the surface points of the sphere by the rigid translation $\mathbf{U}$ and the rigid rotation $\boldsymbol{\Omega}$. For each point $\mathbf{r}\in\partial V$, these displacements must be identical to the displacements of the matrix stuck to the sphere surface. The total matrix displacement on the surface is given on the right-hand side. There, the first term, i.e., the integral, includes all contributions to the matrix displacements due to the surface force density $\mathbf{f}(\mathbf{r'})$ exerted by the particle onto the matrix. The second term, i.e.\ $\mathbf{u}(\mathbf{r})$, corresponds to the (externally) imposed deformation field. At this point, one may be concerned with the validity of the equation, as the Green's function $\mathbf{\hspace{.02cm}\underline{\hspace{-.02cm}G}}$ was derived for an infinitely extended matrix. This seems to contradict the presence of a finite-sized rigid embedded sphere. However, for our calculation it is irrelevant whether we consider the sphere to be rigid inside, or whether it is filled with deformable elastic matrix material as well. The only important point is that the surface shell, which may be considered as infinitely thin, is rigidly translated and rotated as one rigid object. 

Integration of both sides of Eq.~(\ref{faxen_grund}) over $\partial V$ gives
\begin{equation}
	4\pi a^2 U_i ={} \int_{\partial V} \int_{\partial V} G_{ij}(\mathbf{r}-\mathbf{r}^\prime)f_j(\mathbf{r}^\prime) \mathrm{d}S^\prime\mathrm{d}S  + \int_{\partial V} u_i(\mathbf{r})\mathrm{d}S \label{faxen_grund_f}.
\end{equation}
Using Eq.~(\ref{sphere_relation}), the first term on the right-hand side can be connected to the displacement of the sphere due to an external force $\mathbf{F}$.
On $\partial V$, the resulting expression is further simplified using Eqs.~(\ref{stokes}) and (\ref{displacement_sphere}).

For the evaluation of the second term on the right-hand side, we insert the Taylor expansion of $u_i(\mathbf{r})$ 
around the particle center at $\mathbf{r}=\mathbf{r}_0$,
\begin{widetext}
\begin{eqnarray}
	u_i(\mathbf{r}) &={} &u_i(\mathbf{r}_0) + (\mathbf{r}-\mathbf{r}_0)_j \Big[ \nabla_j u_i(\mathbf{r}) \Big]_{\mathbf{r}=\mathbf{r}_0}+ \frac{1}{2}(\mathbf{r}-\mathbf{r}_0)_j(\mathbf{r}-\mathbf{r}_0)_k \Big[ \nabla_j\nabla_k u_i(\mathbf{r})\Big]_{\mathbf{r}=\mathbf{r}_0} \notag\\
	& {} &
	 +\frac{1}{3!}(\mathbf{r}-\mathbf{r}_0)_j(\mathbf{r}-\mathbf{r}_0)_k(\mathbf{r}-\mathbf{r}_0)_l \Big[ \nabla_j\nabla_k\nabla_l u_i(\mathbf{r})\Big]_{\mathbf{r}=\mathbf{r}_0} + ... \label{taylor_u}
\end{eqnarray}\end{widetext}

\noindent Since there are no body forces generating the imposed field $\mathbf{u}(\mathbf{r})$ at $\mathbf{r}=\mathbf{r}_0$, Eq.~(\ref{biharmonic}) must hold, i.e.\ $\nabla^4\mathbf{u}(\mathbf{r}=\mathbf{r}_0)=\mathbf{0}$.
Thus, under the integral, terms of fourth and higher even order in $\nabla$ must vanish due to isotropy.
Furthermore, all odd terms in $(\mathbf{r}-\mathbf{r}_0)$ of the Taylor series must vanish during integration due to symmetry. 
Taking this into account, the second term on the right-hand side of Eq.~(\ref{faxen_grund_f}) can be evaluated as
\begin{widetext}
\begin{eqnarray}
	\int_{\partial V} u_i(\mathbf{r})\,\mathrm{d}S &={} &4\pi a^2 u_i(\mathbf{r}_0) + \frac{1}{2}\int_{\partial V}(\mathbf{r}-\mathbf{r}_0)_j (\mathbf{r}-\mathbf{r}_0)_k  \Big[\nabla_j\nabla_k u_i(\mathbf{r})\Big]_{\mathbf{r}=\mathbf{r}_0} \mathrm{d}S \notag\\
	&={} &4\pi a^2\left(1+\frac{a^2}{6}\nabla^2\right)u_i(\mathbf{r})\bigg|_{\mathbf{r}=\mathbf{r}_0}.
\end{eqnarray}
\end{widetext}
Here, in the step from the first to the second line, we have used that
\begin{equation}\label{delta_1}
	\int_{\partial V} r_j r_k \,\mathrm{d}S ={} \frac{4\pi a^4}{3}\delta_{jk}.
\end{equation}

Collecting all results, Eq.~(\ref{faxen_grund_f}) leads to
\begin{equation}\label{faxen_v}
	 \mathbf{U} ={} \frac{5-6\nu}{24\pi(1-\nu)\mu a}\mathbf{F} + \left(1+\frac{a^2}{6}\nabla^2 \right)\mathbf{u}(\mathbf{r})\bigg|_{\mathbf{r}=\mathbf{r}_0}.
\end{equation}
In this expression, the first contribution to the rigid translation is caused by the external force $\mathbf{F}$, see our previous result in Eq.~(\ref{stokes}). 
The second contribution is due to the imposed matrix displacement field $\mathbf{u}(\mathbf{r})$. As we can see, the sphere is not simply advected by the imposed displacement. Due to its finite size, the additional contribution $\frac{a^2}{6}\nabla^2$ arises. 

In the absence of an external force on the sphere, i.e., for $\mathbf{F}=\mathbf{0}$, we obtain what is referred to as Fax\'en's first law in hydrodynamics \cite{batchelor1972hydrodynamic}:
\begin{equation}
\mathbf{U}^\text{Fax\'en} ={} \left(1+\frac{a^2}{6}\nabla^2\right)\mathbf{u}(\mathbf{r})\bigg|_{\mathbf{r}=\mathbf{r}_0}.\label{faxen_f}
\end{equation}
This relation describes the rigid translation of a rigid sphere in an imposed deformation of the surrounding matrix.

To obtain corresponding expressions for the rotation vector and for the stresslet, we multiply both sides of Eq.~(\ref{faxen_grund}) with $(\mathbf{r}-\mathbf{r}_0)_k$ and integrate over $\partial V$,
\begin{widetext}
\begin{equation} 
\int_{\partial V} (\mathbf{r}-\mathbf{r}_0)_k[\boldsymbol{\Omega}\times(\mathbf{r}-\mathbf{r}_0)]_i \,\mathrm{d}S ={}  \int_{\partial V}\int_{\partial V} (\mathbf{r}-\mathbf{r}_0)_k\, G_{ij}(\mathbf{r}-\mathbf{r}^\prime)f_j(\mathbf{r}^\prime) \,\mathrm{d}S\mathrm{d}S^\prime +\int_{\partial V}(\mathbf{r}-\mathbf{r}_0)_k u_i(\mathbf{r})\,\mathrm{d}S.\label{faxen_grund_D}
\end{equation}\end{widetext}
The integral on the left-hand side is easily evaluated using Eq.~(\ref{delta_1}) and reads
\begin{equation}\label{epsilon_ilk_Omega_l}
	\frac{4\pi a^4}{3}\epsilon_{ilk}\Omega_l.
\end{equation}
In order to calculate the inner integral of the first term on the right-hand side, we substitute $\mathbf{r}^{\prime\prime}=\mathbf{r}-\mathbf{r}_0$ and express the integral in terms of the Fourier transform of the Green's function,
\begin{eqnarray}
\lefteqn{
	\int_{\partial V} G_{ij}(\mathbf{r}-\mathbf{r}^\prime)\,(\mathbf{r}-\mathbf{r}_0)_k \,\mathrm{d}S}\notag\\
	 &={} &\int_{\partial V} G_{ij}(\mathbf{r}^{\prime\prime}-\mathbf{r}^\prime+\mathbf{r}_0)r_k^{\prime\prime} \,\mathrm{d}S^{\prime\prime}\notag\\
	&={} &\frac{1}{(2\pi)^3}\int_{\partial V} \mathrm{d}S^{\prime\prime} \int \mathrm{d}^3k\, \tilde{G}_{ij}(\mathbf{k})r_k^{\prime\prime}e^{i \mathbf{k}\cdot(\mathbf{r}^{\prime\prime}-\mathbf{r}^\prime+\mathbf{r}_0)}.\quad\label{a41}
\end{eqnarray}
Now the integral with respect to $\mathbf{r}^{\prime\prime}$ can be evaluated as
\begin{eqnarray}
	\int_{\partial V}\mathrm{d}S^{\prime\prime}e^{i\mathbf{k}\cdot\mathbf{r}^{\prime\prime}}r_k^{\prime\prime} &={} &-i\nabla_{\mathbf{k}, k}\int_{\partial V}\mathrm{d}S^{\prime\prime}e^{i\mathbf{k}\cdot\mathbf{r}^{\prime\prime}}\notag\\
	&={} &-4\pi i a^2 \hat{k}_k\frac{\mathrm{d}}{\mathrm{d}k}\frac{\sin(ka)}{ka}.
\end{eqnarray}
The integral $\int \mathrm{d}^3 k$ in Eq.~(\ref{a41}) can be split into $\int \mathrm{d}S(\mathbf{\hat{k}})\int_0^\infty k^2\mathrm{d}k$. Inserting Eq.~(\ref{green_fourier}), Eq.~(\ref{a41}) becomes
\begin{widetext}
\begin{eqnarray}
	\lefteqn{\frac{2\pi^2\mu}{a^2}\int_{\partial V}G_{ij}(\mathbf{r}-\mathbf{r}^\prime)\,(\mathbf{r}-\mathbf{r}_0)_k\, \mathrm{d}S}\notag \\
	&={} &-i\int_{\partial V}\mathrm{d}S(\mathbf{\hat{k}})\left(\delta_{ij}-\frac{1}{2(1-\nu)}\hat{k}_i\hat{k}_j\right)\hat{k}_k\int\limits_{0}^{\infty}\mathrm{d}k\, e^{ik\mathbf{\hat{k}}\cdot(\mathbf{r}_0-\mathbf{r}^\prime)}\frac{\mathrm{d}}{\mathrm{d}k}\frac{\sin(ka)}{ka}\notag\\
	&={} &-i \int_{\partial V}\mathrm{d}S(\mathbf{\hat{k}})\left(\delta_{ij}-\frac{1}{2(1-\nu)}\hat{k}_i\hat{k}_j\right)\hat{k}_k
	\bigg(\frac{\sin(ka)}{ka}e^{ik\mathbf{\hat{k}}\cdot(\mathbf{r}_0-\mathbf{r}^\prime)}\bigg|_0^\infty - i \hat{k}_l(\mathbf{r}_0 -\mathbf{r}^\prime)_l \int\limits_{0}^{\infty}\mathrm{d}k\,\frac{\sin(ka)}{ka} e^{ik\mathbf{\hat{k}}\cdot(\mathbf{r}_0-\mathbf{r}^\prime)}\bigg)
\notag\\
&=&{} \int_{\partial V}\mathrm{d}S(\mathbf{\hat{k}})\left(\delta_{ij}-\frac{1}{2(1-\nu)}\hat{k}_i\hat{k}_j\right)\hat{k}_k
	 \hat{k}_l(\mathbf{r}^\prime-\mathbf{r}_0)_l  \int\limits_{0}^{\infty}\mathrm{d}k\,\frac{\sin(ka)}{ka} e^{ik\mathbf{\hat{k}}\cdot(\mathbf{r}_0-\mathbf{r}^\prime)}.
\end{eqnarray}
In the last line, the imaginary part is odd in $\mathbf{\hat{k}}$ and therefore vanishes upon integration.
The remaining real part is an even function in both $\mathbf{\hat{k}}$ and $k$, so that, under the $\int \mathrm{d}S(\mathbf{\hat{k}})$-integral, we may rewrite the $\int\mathrm{d}k$-integral as
\begin{equation}\label{SIN_KA}
	\frac{1}{2}\int\limits_{-\infty}^{\infty}\mathrm{d}k\,\frac{\sin(ka)}{ka}e^{ik\mathbf{\hat{k}}\cdot(\mathbf{r}_0-\mathbf{r}^\prime)} ={} \begin{cases}
	\frac{\pi}{2a}, &\text{for }-1<\frac{\mathbf{\hat{k}}\cdot(\mathbf{r}^\prime-\mathbf{r}_0)}{a}<1,\\
	0, &\text{otherwise},
	\end{cases}
\end{equation}
see Appendix B.
We obtain
\begin{equation}
	\int_{\partial V}G_{ij}(\mathbf{r}-\mathbf{r}^\prime)\,(\mathbf{r}-\mathbf{r}_0)_k \,\mathrm{d}S
	={} \frac{a}{4\pi\mu}(\mathbf{r}^\prime-\mathbf{r}_0)_l\int_{\Delta S}\mathrm{d}S(\mathbf{\hat{k}})\left(\delta_{ij}-\frac{1}{2(1-\nu)}\hat{k}_i\hat{k}_j\right)\hat{k}_k\hat{k}_l,
\end{equation}
where the surface of integration $\Delta S$ is given by
\begin{equation}
	\Delta S = \left\{ \mathbf{\hat{k}} \,\,\bigg| -1 < \frac{\mathbf{\hat{k}}\cdot(\mathbf{r}^\prime-\mathbf{r}_0)}{a} < 1 \right\}.
\end{equation}
Since $\mathbf{r}^\prime$ is located on the surface of the inclusion, i.e.\ $|\mathbf{r}^\prime-\mathbf{r}_0|=a$, $\Delta S$ corresponds to the surface of the unit sphere.
Using Eq.~(\ref{delta_1}) (for $\mathbf{\hat{k}}$ instead of $\mathbf{r}$) and
\begin{equation}
	\int_{\Delta S}\hat{k}_i \hat{k}_j \hat{k}_k \hat{k}_l \,\mathrm{d}S(\mathbf{\hat{k}}) ={} \frac{4\pi}{15}(\delta_{ij}\delta_{kl} + \delta_{ik}\delta_{jl} + \delta_{il}\delta_{jk})\label{delta_2}
\end{equation}
finally leads to
\begin{eqnarray}
\lefteqn{\int_{\partial V} (\mathbf{r}-\mathbf{r}_0)_k G_{ij}(\mathbf{r}-\mathbf{r}^\prime)f_j(\mathbf{r}^\prime) \,\mathrm{d}S}\notag\\
&={} &\frac{a}{15\mu}\Bigg[5(\mathbf{r}^\prime-\mathbf{r}_0)_k f_i - \frac{1}{2(1-\nu)}\big((\mathbf{r}^\prime-\mathbf{r}_0)_k f_i
 +(\mathbf{r}^\prime-\mathbf{r}_0)_i f_k+(\mathbf{r}^\prime-\mathbf{r}_0)_l f_l \delta_{ik}\big)\Bigg].\label{r_k_G_ij_f_j_dS}
\end{eqnarray}

The second term on the right-hand side of Eq.~(\ref{faxen_grund_D}) can be evaluated by inserting the Taylor expansion of $\mathbf{u}(\mathbf{r})$ from Eq.~(\ref{taylor_u}),
\begin{eqnarray}\label{r_k_u_i_dS}
	\int_{\partial V} (\mathbf{r}-\mathbf{r}_0)_k u_i(\mathbf{r})\,\mathrm{d}S
	 &={} &\int_{\partial V} (\mathbf{r}-\mathbf{r}_0)_k(\mathbf{r}-\mathbf{r}_0)_j\Big[\nabla_j u_i(\mathbf{r})\Big]_{\mathbf{r}=\mathbf{r}_0}\mathrm{d}S \notag\\ &{}&+\frac{1}{6}\int_{\partial V}(\mathbf{r}-\mathbf{r}_0)_k (\mathbf{r}-\mathbf{r}_0)_j (\mathbf{r}-\mathbf{r}_0)_l ( \mathbf{r}-\mathbf{r}_0)_m \Big[\nabla_j \nabla_l \nabla_m u_i(\mathbf{r})\Big]_{\mathbf{r}=\mathbf{r}_0}\mathrm{d}S \notag\\
	&={} &\frac{4\pi a^4}{3}\left(1+\frac{a^2}{10}\nabla^2\right)\nabla_k u_i(\mathbf{r})\bigg|_{\mathbf{r}=\mathbf{r}_0},
\end{eqnarray}
where again we have used Eqs.~(\ref{delta_1}) and (\ref{delta_2}) (for $(\mathbf{r}-\mathbf{r}_0)$ instead of $\mathbf{\hat{k}}$ in the latter).
The other terms in the expansion again vanish due to isotropy and symmetry upon integration.

Altogether, combining Eqs.~(\ref{faxen_grund_D}), (\ref{epsilon_ilk_Omega_l}), (\ref{r_k_G_ij_f_j_dS}), and (\ref{r_k_u_i_dS}), we find
\begin{eqnarray}
	\frac{4\pi a^4}{3}\epsilon_{ilk}\Omega_l &={} &\frac{a}{15\mu}\int_{\partial V} \mathrm{d}S^\prime \Bigg[5(\mathbf{r}^\prime-\mathbf{r}_0)_k f_i - \frac{1}{2(1-\nu)}\Big((\mathbf{r}^\prime-\mathbf{r}_0)_k f_i +(\mathbf{r}^\prime-\mathbf{r}_0)_i f_k+(\mathbf{r}^\prime-\mathbf{r}_0)_l f_l \delta_{ik}\Big)\Bigg] \notag\\
	&{} &
	+\frac{4\pi a^4}{3}\left(1+\frac{a^2}{10}\nabla^2\right)\nabla_k u_i(\mathbf{r})\bigg|_{\mathbf{r}=\mathbf{r}_0}.\label{grund_D2}
\end{eqnarray}\end{widetext}

This tensor equation can be split into a symmetric and an antisymmetric part. First, we calculate the antisymmetric part by multiplying Eq.~(\ref{grund_D2}) by $\epsilon_{ijk}$. Since there are no body forces generating the imposed field $\mathbf{u}(\mathbf{r})$ at $\mathbf{r}=\mathbf{r}_0$, Eq.~(\ref{rot_u}) most hold for the last term, i.e.\ $\nabla\times\nabla^2\mathbf{u}(\mathbf{r}=\mathbf{r}_0)=\mathbf{0}$. Therefore the $\frac{a^2}{10}\nabla^2$-term in Eq.~(\ref{grund_D2}) vanishes.
Using the definition of the torque from Eq.~(\ref{torque}), we obtain
\begin{equation}\label{faxen_w}
	\boldsymbol{\Omega} ={} \frac{1}{8\pi\mu a^3}\mathbf{T} + \frac{1}{2}\nabla\times \mathbf{u}(\mathbf{r})\bigg|_{\mathbf{r}=\mathbf{r}_0}.
\end{equation}
$\mathbf{T}$ corresponds to an external torque acting onto the sphere, which is transmitted by the sphere onto the surrounding matrix (with the reference point of the torque at the center of the sphere). 

Similarly to the previous case of rigid translations, in the absence of an external torque acting on the sphere, i.e., for $\mathbf{T}=\mathbf{0}$, we obtain a relation referred to as Fax\'en's second law in hydrodynamics \cite{batchelor1972hydrodynamic}: 
\begin{equation}
		\boldsymbol{\Omega}^\text{Fax\'en} ={} \frac{1}{2}\nabla\times\mathbf{u}(\mathbf{r})\bigg|_{\mathbf{r}=\mathbf{r}_0}\label{faxen_r}. 
\end{equation}
This relation quantifies the (static) rigid rotation of a rigid sphere in an imposed deformation of the surrounding matrix.

Finally, we calculate the symmetric part of Eq.~(\ref{grund_D2}). The $\boldsymbol{\Omega}$-term vanishes because of its antisymmetry. Thus, we find
\begin{widetext}
\begin{eqnarray}
	0
	&={}&\frac{a}{15\mu}\frac{1}{2(1-\nu)}\int_{\partial V} \mathrm{d}S^\prime \bigg[(4-5\nu)\Big((\mathbf{r}^\prime-\mathbf{r}_0)_i f_k + (\mathbf{r}^\prime-\mathbf{r}_0)_k f_i \Big) - (\mathbf{r}^\prime-\mathbf{r}_0)_jf_j\delta_{ik}\bigg]\notag\\
	 &{}&+ \frac{4\pi a^4}{3}\left(1+\frac{a^2}{10}\nabla^2\right)\frac{1}{2}\big(\nabla_i u_k(\mathbf{r})+ \nabla_k u_i(\mathbf{r})\big)\bigg|_{\mathbf{r}=\mathbf{r}_0}\notag\\
	 &=:{} &\frac{1}{2}(A_{ik}+A_{ki}).
\end{eqnarray}
To obtain an expression solely for the stresslet as defined in Eq.~(\ref{stresslet}), we add a vanishing trace term
\begin{equation}
	\frac{1}{5(1-2\nu)}A_{jj}\delta_{ik} ={}  \frac{a}{15\mu} \frac{1}{2(1-\nu)}\int_{\partial V} \mathrm{d}S^\prime (\mathbf{r}^\prime - \mathbf{r}_0)_j f_j\delta_{ik}  + \frac{4\pi a^4}{15}\left( 1+\frac{a^2}{10}\nabla^2 \right)\frac{1}{1-2\nu}\nabla_j u_j(\mathbf{r})\delta_{ik}\bigg|_{\mathbf{r}=\mathbf{r}_0},
\end{equation}
leading to 
\begin{equation}\label{stresslet_tensor_eq}
	0={} \frac{1}{2}(A_{ik}+A_{ki})+\frac{1}{5(1-2\nu)}A_{jj}\delta_{ik}.
\end{equation}
Then, the definition of $S_{ik}$ appears in Eq.~(\ref{stresslet_tensor_eq}). Solving for $S_{ik}$, we find the stresslet as
\begin{equation}
	\mathbf{\underline{S}} ={} -\frac{4\pi(1-\nu)\mu a^3}{4-5\nu}\left(1+\frac{a^2}{10}\nabla^2\right)\Bigg[\frac{1}{1-2\nu}\mathbf{\underline{\hat{I}}}\nabla\cdot\mathbf{u}(\mathbf{r}) +\frac{5}{2}\Big(\nabla\mathbf{u}(\mathbf{r})+\big(\nabla\mathbf{u}(\mathbf{r})\big)^T\Big) \Bigg]\Bigg|_{\mathbf{r}=\mathbf{r}_0}, \label{faxen_3}
\end{equation}\end{widetext}
\mpu{where the superscript $(\bullet)^T$ marks the transpose.}

Eq.~(\ref{faxen_3}) expresses the stress that a rigid spherical inclusion exerts onto the surrounding matrix in the imposed displacement field $\mathbf{u}(\mathbf{r})$ of the matrix. The matrix deformation is imposed from elsewhere, that is, not by the spherical inclusion itself. However, the inclusion due to its rigidity resists this deformation. This resistance leads to the described stresslet. 
 
Vice versa, the stresslet that the matrix exerts onto the particle is given by
\begin{equation}\label{stresslet_deformable}
	\mathbf{\underline{S}}^\text{Fax\'en}={} -\mathbf{\underline{S}},
\end{equation}
which together with Eq.~(\ref{faxen_3}) may be referred to as Fax\'en's third law and was derived by Batchelor in the hydrodynamic case \cite{batchelor1972hydrodynamic}.

\section{Displaceability and rotateability matrix}\label{Section_7_displaceability_rotateability}
Now we have all the ingredients to consider the coupled displacements and rotations of $N$ spherical inclusions embedded in the infinitely extended homogeneous elastic medium. 
For simplicity, we consider identical spheres of radius $a$, labeled by $1,...,N$. 

We here adhere to the following cause-and-effect chain. 
Each spherical inclusion $j$ is subject to an external force $\mathbf{F}_j$ and an external torque $\mathbf{T}_j$, $j=1,...,N$.
As a consequence of these forces and torques, the inclusions are displaced and rotated by rigid translation vectors $\mathbf{U}_i$ and rigid rotation vectors $\boldsymbol{\Omega}_i$, respectively, $i=1,...,N$. \mpu{Moreover}, the spheres transmit the forces and torques to the surrounding elastic medium, causing additional deformations in their environment. Other inclusions are exposed to these induced deformations and counteract due to their rigidity. This leads to further distortions, acting back on all other rigid spheres that likewise resist induced deformations, resulting in mutually coupled particle translations and rotations. In the following, we derive analytical expressions for these translations and rotations, using the external forces and torques as an input.

In formal analogy to the hydrodynamic mobility matrices \cite{dhont1996introduction,reichert2004hydrodynamic}, we can define \textit{elastic displaceability and rotateability matrices}. 
%
Given the external (quasi)static forces $\mathbf{F}_j$ and (quasi)static torques $\mathbf{T}_j$, $j=1,...,N$, applied to the spherical inclusions, these matrices directly express the caused displacements $\mathbf{U}_i$ and rotations $\boldsymbol{\Omega}_i$ in the resulting situation of new (quasi)static equilibrium, $i=1,...,N$:
\begin{widetext}
\begin{eqnarray}\label{displaceability_matrix}
\begin{pmatrix}
\mathbf{U}_1\\
\vdots\\
\mathbf{U}_N\\[1em]
\boldsymbol{\Omega}_1\\
\vdots\\
\boldsymbol{\Omega}_N
\end{pmatrix}
={}
\begin{pmatrix}
\mathbf{\underline{M}}_{11}^\text{tt} & \cdots & \mathbf{\underline{M}}_{1N}^\text{tt} & \mathbf{\underline{M}}_{11}^\text{tr} &  \cdots & \mathbf{\underline{M}}_{1N}^\text{tr} \\
\vdots & \ddots & \vdots &\vdots & \ddots & \vdots \\
\mathbf{\underline{M}}_{N1}^\text{tt} & \cdots & \mathbf{\underline{M}}_{NN}^\text{tt} &
\mathbf{\underline{M}}_{N1}^\text{tr} & \cdots & \mathbf{\underline{M}}_{NN}^\text{tr}\\[1em]
\mathbf{\underline{M}}_{11}^\text{rt} &  \cdots & \mathbf{\underline{M}}_{1N}^\text{rt} & \mathbf{\underline{M}}_{11}^\text{rr} & \cdots & \mathbf{\underline{M}}_{1N}^\text{rr} \\
\vdots & \ddots & \vdots &\vdots & \ddots & \vdots \\
\mathbf{\underline{M}}_{N1}^\text{rt} & \cdots & \mathbf{\underline{M}}_{NN}^\text{rt}
&\mathbf{\underline{M}}_{N1}^\text{rr} & \cdots & \mathbf{\underline{M}}_{NN}^\text{rr}

\end{pmatrix}
\cdot
\begin{pmatrix}
\mathbf{F}_1\\
\vdots\\
\mathbf{F}_N\\[1em]
\mathbf{T}_1\\
\vdots\\
\mathbf{T}_N
\end{pmatrix}
\end{eqnarray}
%
\end{widetext}

Here, the sub-matrices $\mathbf{\underline{M}}_{ij}^\text{tt}$ express how the particles are translated due to the forces acting on all the particles (translation--translation coupling, $i,j=1,...,N$). Their components have been derived already in a previous work \cite{puljiz2016forces}. The sub-matrices $\mathbf{\underline{M}}_{ij}^\text{tr}$ include contributions to the translations due to the torques acting on the inclusions (translation--rotation coupling). Similarly, the sub-matrices $\mathbf{\underline{M}}_{ij}^\text{rt}$ determine how forces acting on the particles lead to their rotations (rotation--translation coupling). The cause of rotations by torques is given by the sub-matrices $\mathbf{\underline{M}}_{ij}^\text{rr}$ (rotation--rotation coupling). 

We stress that the role of the surrounding elastic medium is implicitly contained in these matrices. Their components will solely depend on the configuration of the rigid inclusions. Therefore, they significantly facilitate the problem of calculating the coupled displacements and rotations described above. It is not necessary any longer to explicitly calculate the displacement field $\mathbf{u}(\mathbf{r})$ of the surrounding medium once the expressions for these matrices have been derived.

Below, we shall explicitly perfom this derivation for the components $\mathbf{\underline{M}}_{ij}^\text{tt}$, $\mathbf{\underline{M}}_{ij}^\text{tr}$, $\mathbf{\underline{M}}_{ij}^\text{rt}$, and $\mathbf{\underline{M}}_{ij}^\text{rr}$ as an expansion in the inverse separation distance of the inclusions. Here, we proceed up to (including) fourth order. This comprises pairwise interactions mediated by the surrounding elastic medium, see Sec.~\ref{Section_8_two_rigid_sphere_interaction}, and three-body interactions, see Sec.~\ref{Section_9_three_spheres}.

\section{Two-body interactions}\label{Section_8_two_rigid_sphere_interaction}

In the following, we start from the forces and torques acting on the inclusions, which as a consequence \mpu{leads} to the coupled particle translations and rotations. Our approach adapts the \textit{method of reflections} from the hydrodynamic literature as presented in Ref.~\onlinecite{dhont1996introduction}. In addition to that, we here explicitly include the role of imposed torques as for instance exerted by external magnetic fields on magnetically anisotropic inclusions. 
Moreover, we take into account the rigidity of the inclusions directly via the stresslets that follow from their resistance to deformations \cite{batchelor1972hydrodynamic,karrila1991microhydrodynamics}. 

The initial forces and torques acting on the inclusions are either imposed externally, or they are induced between the inclusions from outside. These are \textit{not} the forces and torques exerted by the elastic matrix onto the inclusions. For clarity, we consider the influence of the imposed or induced forces and torques separately in two steps. Due to the linearity of the governing equations, the results of these two steps can in the end simply be added/superimposed.

\subsection{Forces imposed on or induced between the inclusions}\label{subsection_forces_imposed}

\begin{figure}
\centerline{\includegraphics[width=\columnwidth]{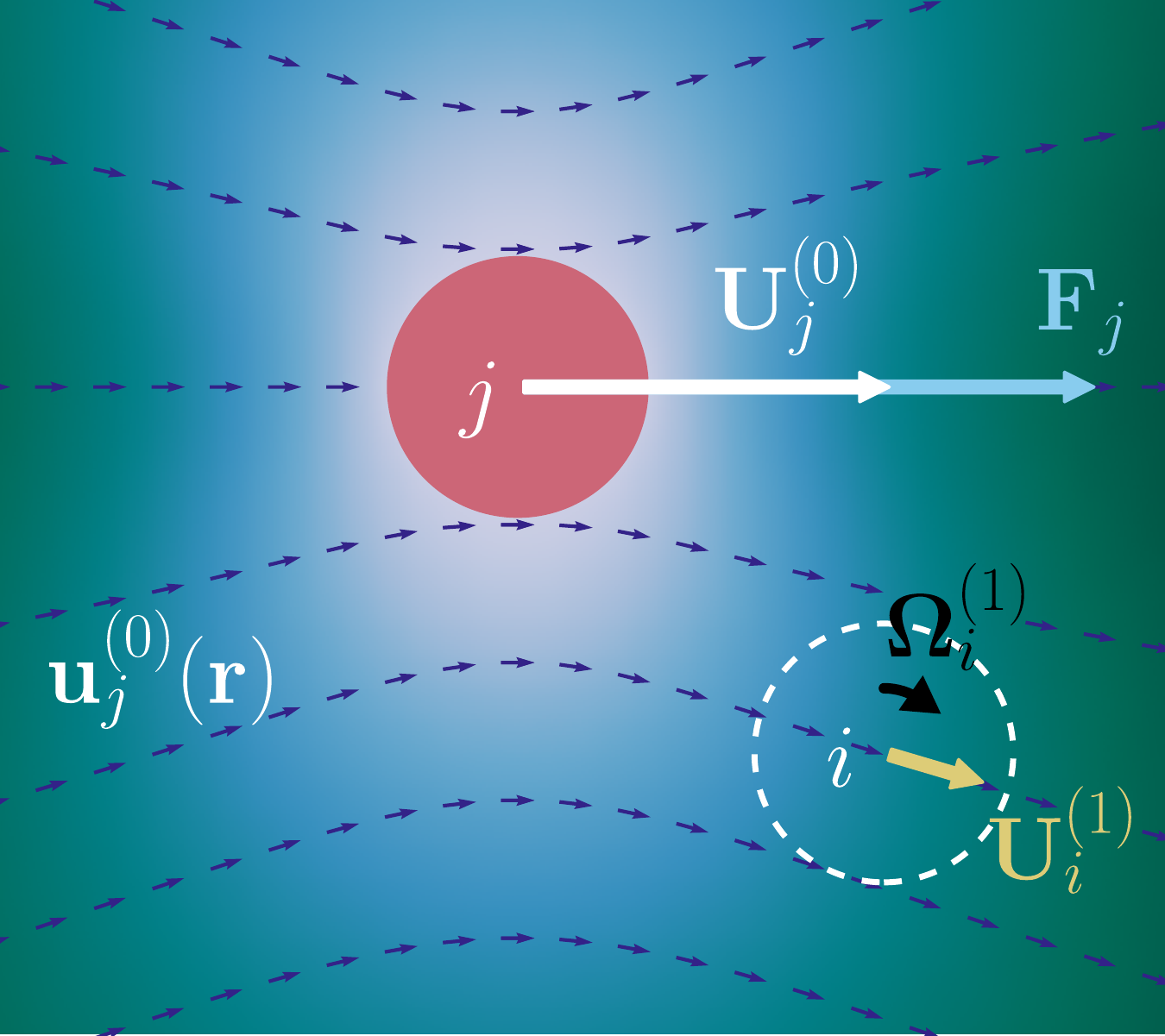}}
\caption{
Illustration of the immediate effect that the displacement of sphere $j$ has on the translation and rotation of another sphere $i$.
A force $\mathbf{F}_j$ is externally imposed on sphere $j$. As a consequence, sphere $j$ gets rigidly translated as given by $\mathbf{U}_j^{(0)}$, see Eq.~(\ref{v_j_0}).
Moreover, the surrounding matrix is distorted, as described by the displacement field $\mathbf{u}_j^{(0)}(\mathbf{r})$, see Eq.~(\ref{u_2_0}).
The local directions of $\mathbf{u}_j^{(0)}(\mathbf{r})$ are indicated by the small arrows that, for visibility, are rescaled to identical length. We indicated the local magnitude of $\mathbf{u}_j^{(0)}(\mathbf{r})$ by background color, where brighter color represents higher magnitude and the color values follow an arc-tangent scale.
Sphere $i$ is exposed to the induced displacement field $\mathbf{u}_j^{(0)}(\mathbf{r})$ and therefore gets translated as denoted by $\mathbf{U}_i^{(1)}$ and rotated as denoted by $\boldsymbol{\Omega}_i^{(1)}$. These quantities can be calculated from $\mathbf{u}_j^{(0)}(\mathbf{r})$ via Eqs.~(\ref{v_1_1_f}) and (\ref{w_1_1_f}), respectively, leading to Eqs.~(\ref{v1_1_explicit}) and (\ref{w1_1_explicit}).
Overall, in this way we obtain the corresponding contributions to the \textit{displaceability and rotateability matrices} $\mathbf{\underline{M}}_{i=j}^\text{tt}$, $\mathbf{\underline{M}}_{i\neq j}^\text{tt}$, $\mathbf{\underline{M}}_{i=j}^\text{rt}$, and $\mathbf{\underline{M}}_{i\neq j}^\text{rt}$ in Eqs.~(\ref{M_II_2}), (\ref{M_IJ_2}), (\ref{M_II_rt}), and (\ref{M_IJ_rt}), respectively, up to inverse quartic order in the particle distances.}
\label{fig1}
\end{figure}

\begin{figure*}
\centerline{\includegraphics[width=\columnwidth]{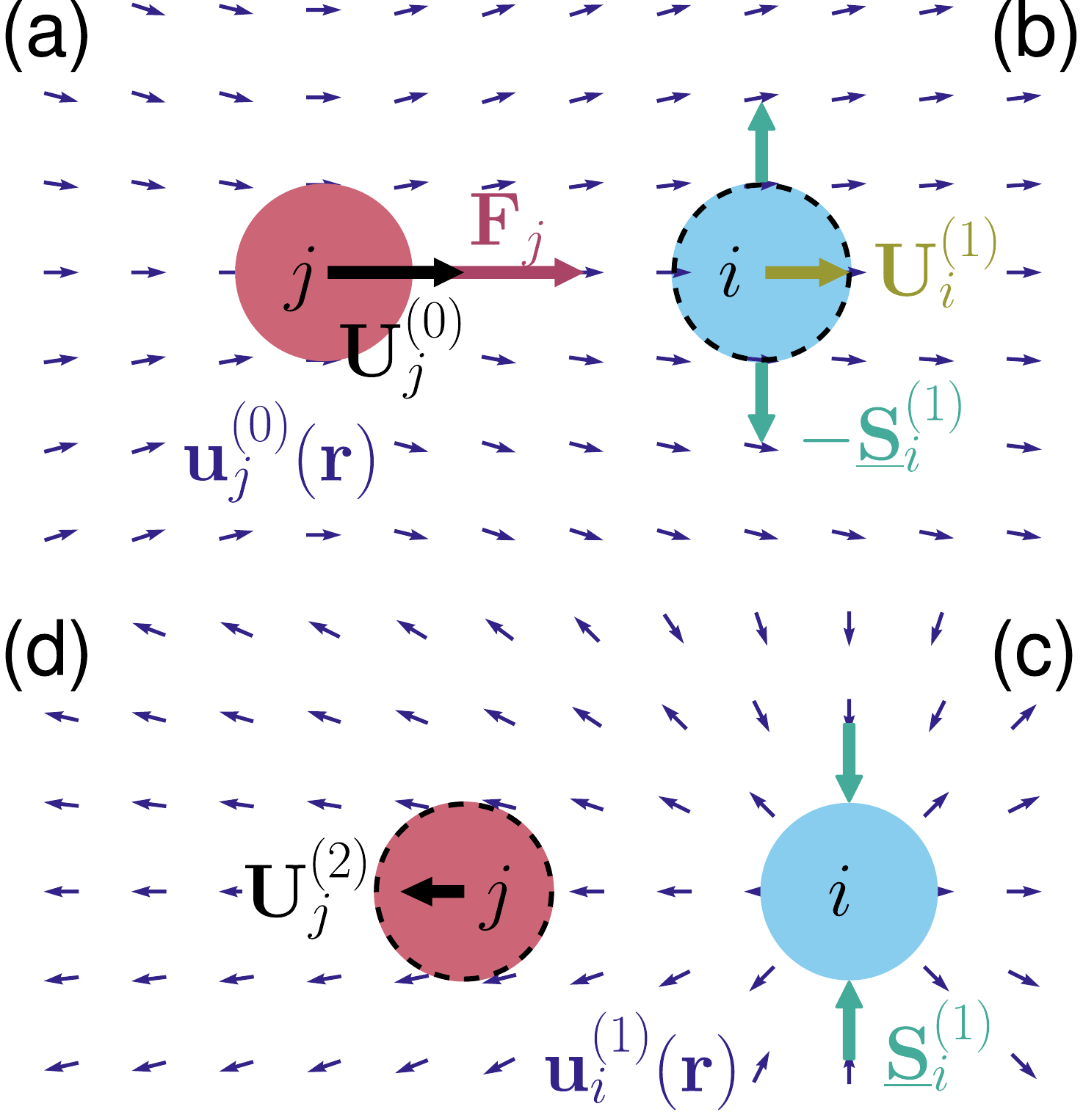}}
\caption{ 
Illustration of the rigidity-based \textit{reflection} of an induced displacement field by another sphere.
(a) As in Fig.~\ref{fig1}, an externally imposed force $\mathbf{F}_j$ acts onto the spherical particle $j$. This directly results in the particle translation $\mathbf{U}_j^{(0)}$ and in the displacement field $\mathbf{u}^{(0)}_j(\mathbf{r})$ in the surrounding elastic matrix, see Eqs.~(\ref{v_j_0}), (\ref{u_2_0}), and Fig.~\ref{fig1}. The small arrows indicate the local direction of the induced displacement fields.
(b) Particle $i$ is exposed to the displacement field $\mathbf{u}^{(0)}_j(\mathbf{r})$ and is therefore translated by $\mathbf{U}^{(1)}_i$, see Eq.~(\ref{v_1_1_f}). Rotations are not considered here for simplicity. Simultaneously, the displacement field tends to deform particle $i$ as given by the stresslet $-\mathbf{\underline{S}}_i^{(1)}$, see Eqs.~(\ref{faxen_3}) and (\ref{stresslet_deformable}).
(c) However, the rigid particle $i$ resists deformation and imposes the stresslet $\mathbf{\underline{S}}_i^{(1)}$ onto the surrounding elastic matrix, see Eq.~(\ref{2s_stress}). $\mathbf{\underline{S}}_i^{(1)}$ induces yet another displacement field $\mathbf{u}_i^{(1)}(\mathbf{r})$ in the elastic environment, see Eq.~(\ref{u_i_1}), which overlays the initial field $\mathbf{u}_j^{(0)}(\mathbf{r})$. In this way, the initial field $\mathbf{u}^{(0)}_j(\mathbf{r})$ gets partially \textit{reflected} by the rigid particle $i$, leading to $\mathbf{u}^{(1)}_i(\mathbf{r})$.
(d) Now, particle $j$ is exposed to $\mathbf{u}^{(1)}_i(\mathbf{r})$. Its initial translation $\mathbf{U}^{(0)}_j$ thus gets corrected by a translation $\mathbf{U}_j^{(2)}$, see Eq.~(\ref{v1_2_explicit}) after swapping indices $i$ and $j$. Altogether, this leads to the quartic contribution in the inverse particle separation distance to the \textit{displaceability matrices} $\mathbf{\underline{M}}_{i=j}^\text{tt}$ in Eq.~(\ref{M_II_2})\mpu{, after} switching $i \leftrightarrow j$. In analogy, we may consider\mpu{,} instead of the initial particle $j$\mpu{,} a different, third particle exposed to the reflected field. Following the same scheme and calculating its induced translation, we obtain the \textit{three-body interaction} included by the contribution $\mathbf{\underline{M}}_{i\neq j}^{\text{tt(3)}}$ in Eq.~(\ref{M_tt_3}). (For the latter purpose, the first, second, and third particle are referred to as $j$, $k$, and $i$, respectively.)
}
\label{fig3}
\end{figure*}

In the following, we consider two rigid spherical inclusions $i$ and $j$, both of radius $a$. They are located at positions $\mathbf{r}_i$ and $\mathbf{r}_j$, respectively. 
The forces $\mathbf{F}_i$ and $\mathbf{F}_j$ are externally applied to the spheres $i$ and $j$, respectively, or induced between them.
As indicated before, we will proceed below by an expansion in the inverse separation distance between the two spheres. 

To zeroth order, the spheres are thus effectively considered to be infinitely far away from each other. Consequently, the interactions between the two spheres via the surrounding elastic matrix do not enter.
The actual translations of the spheres, $\mathbf{U}_i^{(0)}$ and $\mathbf{U}_j^{(0)}$, respectively, are then given by the solution for isolated spherical inclusions, see Eq.~(\ref{stokes}), and read
\begin{eqnarray}
	\mathbf{U}_i^{(0)} &={} &\mathbf{u}_i^{(0)}(\mathbf{r}\in\partial V_i) ={} \frac{5-6\nu}{24\pi(1-\nu)\mu a}\mathbf{F}_i,\label{v_i_0}\\
	\mathbf{U}_j^{(0)} &={} & \mathbf{u}_j^{(0)}(\mathbf{r}\in\partial V_j) ={} \frac{5-6\nu}{24\pi(1-\nu)\mu a}\mathbf{F}_j.\label{v_j_0}
\end{eqnarray}
Furthermore, to zeroth order, the induced displacement field of the elastic matrix around each sphere $i$ and $j$ has been calculated in Eq.~(\ref{displacement_sphere}), i.e.
\begin{eqnarray}
	\mathbf{u}_i^{(0)}(\mathbf{r}) &={} & \left(1+\frac{a^2}{6}\nabla^2\right)\mathbf{\hspace{.02cm}\underline{\hspace{-.02cm}G}}(\mathbf{r}-\mathbf{r}_i)\cdot\mathbf{F}_i,\label{u_1_0}\\
	\mathbf{u}_j^{(0)}(\mathbf{r}) &={} & \left(1+\frac{a^2}{6}\nabla^2\right)\mathbf{\hspace{.02cm}\underline{\hspace{-.02cm}G}}(\mathbf{r}-\mathbf{r}_j)\cdot\mathbf{F}_j.\label{u_2_0}
\end{eqnarray}
In Fig.~\ref{fig1}, $\mathbf{u}_j^{(0)}(\mathbf{r})$ is indicated by the small arrows.

Next, we take into account the mutual interactions between the two spheres mediated by the surrounding elastic matrix.
For example, we consider particle $i$ that is embedded in the elastic matrix. Thus it is exposed to the displacement field $\mathbf{u}_j^{(0)}(\mathbf{r})$ that results from the force $\mathbf{F}_j$ acting on sphere $j$. 
An additional translation $\mathbf{U}^{(1)}_i$ and rotation $\boldsymbol{\Omega}^{(1)}_i$ of sphere $i$ are induced in this way, which we can calculate from the Fax\'en relations, Eqs.~(\ref{faxen_f}) and (\ref{faxen_r}). 
They read
\begin{eqnarray}
	\mathbf{U}_i^{(1)} &={} &\left(1+\frac{a^2}{6}\nabla^2\right)\mathbf{u}_j^{(0)}(\mathbf{r})\bigg|_{\mathbf{r}=\mathbf{r}_i},\label{v_1_1_f}\\
	\boldsymbol{\Omega}_i^{(1)} &={} &\frac{1}{2}\nabla\times\mathbf{u}_j^{(0)}(\mathbf{r})\bigg|_{\mathbf{r}=\mathbf{r}_i}.
\label{w_1_1_f}
\end{eqnarray}
That is, $\mathbf{u}_j^{(0)}(\mathbf{r})$ now plays the role of the imposed matrix displacement field $\mathbf{u}(\mathbf{r})$ in Eqs.~(\ref{faxen_f}) and (\ref{faxen_r}). 

In general, the displacement field $\mathbf{u}_j^{(0)}(\mathbf{r})$ would tend to deform sphere $i$. In other words, a stress is exerted on particle $i$. 
Yet, because of its rigidity, sphere $i$ resists this deformation.
As a consequence, the overall displacement field induced by sphere $j$, i.e.\ $\mathbf{u}_j^{(0)}(\mathbf{r})$, is disturbed via the presence of sphere $i$.
We can find this disturbance from the stress that the rigid sphere $i$ itself exerts back onto the matrix. The corresponding stresslet follows from Eq.~(\ref{faxen_3}) and here takes the form
\begin{widetext}
\begin{equation}
	\mathbf{\underline{S}}_i^{(1)} ={} -\frac{4\pi(1-\nu)\mu a^3}{4-5\nu}\left(1+\frac{a^2}{10}\nabla^2\right)\Bigg[\frac{1}{1-2\nu}\mathbf{\underline{\hat{I}}}\nabla\cdot\mathbf{u}_j^{(0)}(\mathbf{r})+ \frac{5}{2}\Big(\nabla\mathbf{u}_j^{(0)}(\mathbf{r})+\big(\nabla\mathbf{u}_j^{(0)}(\mathbf{r})\big)^T\Big) \Bigg]\Bigg|_{\mathbf{r}=\mathbf{r}_i}.
\label{2s_stress}
\end{equation}
\end{widetext}

Analogous expressions for sphere $j$ are obtained by swapping the indices $i\leftrightarrow j$ in Eqs.~(\ref{v_1_1_f})--(\ref{2s_stress}).

We now proceed to improve our solution by iteration.
For this purpose, we calculate the mentioned disturbances $\mathbf{u}_i^{(1)}(\mathbf{r})$ and $\mathbf{u}_j^{(1)}(\mathbf{r})$ that the stresslets $\mathbf{\underline{S}}_i^{(1)}$ and $\mathbf{\underline{S}}_j^{(1)}$ cause in the matrix, respectively. 
We find corresponding expressions from Eq.~(\ref{multipole_fts}):
\begin{eqnarray}
	\mathbf{u}_i^{(1)}(\mathbf{r}) &={} &-(\mathbf{\underline{S}}_i^{(1)}\cdot\nabla)\cdot\mathbf{\hspace{.02cm}\underline{\hspace{-.02cm}G}}(\mathbf{r}-\mathbf{r}_i), \label{u_i_1}\\
	\mathbf{u}_j^{(1)}(\mathbf{r}) &={} &-(\mathbf{\underline{S}}_j^{(1)}\cdot\nabla)\cdot\mathbf{\hspace{.02cm}\underline{\hspace{-.02cm}G}}(\mathbf{r}-\mathbf{r}_j).\label{u_j_1}
\end{eqnarray}
We should remark that Eq.~(\ref{multipole_fts}) also contains the forces imposed on the inclusions. However, at this stage of iteration, they do not contribute. The direct influence of the forces has already been determined in Eqs.~(\ref{v_i_0})--(\ref{u_2_0}). The spheres simply follow the resulting induced displacement fields, without any additional extra net force or torque resistance, see Eqs.~(\ref{v_1_1_f}) and (\ref{w_1_1_f}). Their only resistance is due to their rigidity as described above, which now enters Eqs.~(\ref{u_i_1}) and (\ref{u_j_1}) in the form of the stresslets. Due to the linearity of the Navier-Cauchy equations, Eq.~(\ref{navier_cauchy}), the disturbances in Eqs.~(\ref{u_i_1}) and (\ref{u_j_1}) can in the end simply be added/superimposed to the displacement fields in Eqs.~(\ref{u_1_0}) and (\ref{u_2_0}). 

In the next step, each sphere is now additionally exposed to one of these rigidity-induced displacement fields $\mathbf{u}_i^{(1)}(\mathbf{r})$ and $\mathbf{u}_j^{(1)}(\mathbf{r})$ created by the other sphere. This leads to yet another contribution to the translation ($\mathbf{U}_i^{(2)}$ and $\mathbf{U}_j^{(2)}$) and rotation ($\boldsymbol{\Omega}_i^{(2)}$ and $\boldsymbol{\Omega}_j^{(2)}$) of each sphere. Again, we can calculate these contributions from the Fax\'en laws, see Eqs.~(\ref{faxen_f}) and (\ref{faxen_r}), now taking $\mathbf{u}_j^{(1)}(\mathbf{r})$ and $\mathbf{u}_i^{(1)}(\mathbf{r})$ as the imposed displacement fields, respectively:
\begin{eqnarray}
	\mathbf{U}_i^{(2)} &={} &\left(1+\frac{a^2}{6}\nabla^2\right)\mathbf{u}_j^{(1)}(\mathbf{r})\bigg|_{\mathbf{r}=\mathbf{r}_i},\label{v1_2_f}\\
	\boldsymbol{\Omega}_i^{(2)} &={} &\frac{1}{2}\nabla\times\mathbf{u}_j^{(1)}(\mathbf{r})\bigg|_{\mathbf{r}=\mathbf{r}_i},\label{w1_2_f}
\end{eqnarray}
with $\mathbf{U}_j^{(2)}$ and $\boldsymbol{\Omega}_j^{(2)}$ obtained by swapping the indices $i\leftrightarrow j$. 
The overall situation resulting in the displacement $\mathbf{U}_j^{(2)}$ is illustrated in Fig.~\ref{fig3} and has already been considered in Ref.~\onlinecite{puljiz2016forces}.

Altogether, one can say that parts of the displacement fields $\mathbf{u}_i^{(0)}(\mathbf{r})$ and $\mathbf{u}_j^{(0)}(\mathbf{r})$, initially generated by the first sphere, are \textit{reflected} by the respectively other sphere in the form of $\mathbf{u}_j^{(1)}(\mathbf{r})$ and $\mathbf{u}_i^{(1)}(\mathbf{r})$. 
This is due to the rigidity of the spheres.
Then these fields are felt again by the corresponding first sphere. 

In principle, one can continue this iteration by considering further reflections. Also the first sphere is rigid and will resist deformations in the reflected field, etc. We can use the same formulae summarized above to continue this iteration. Accordingly, this approach was called \textit{method of reflections} in the hydrodynamic literature \cite{dhont1996introduction}. Overall, it turns out that this iterative procedure corresponds to an expansion in the inverse particle separation distance $r_{ij}^{-1}$, with $r_{ij}=|\mathbf{r}_{i}-\mathbf{r}_{j}|$. Here, we proceed up to (including) the fourth order $r_{ij}^{-4}$. \mpu{Then, counting} factors $r_{ij}^{-1}$ and gradients shows that we may stop at the presented stage.

To find the resulting explicit analytical expressions for the matrix-mediated particle interactions, let us now explicitly calculate the contributions in Eqs.~(\ref{v_1_1_f}), (\ref{w_1_1_f}), (\ref{v1_2_f}), and (\ref{w1_2_f}).
From Eqs.~(\ref{greens_function}), (\ref{u_2_0}), and (\ref{v_1_1_f}), using Eq.~(\ref{biharmonic}), we find for the first correction of the translation of sphere~$i$
\begin{widetext}
 \begin{eqnarray}
 	\mathbf{U}_i^{(1)} &={} &\left(1+\frac{a^2}{3}\nabla^2\right)\mathbf{\hspace{.02cm}\underline{\hspace{-.02cm}G}}(\mathbf{r}-\mathbf{r}_j)\cdot\mathbf{F}_j\bigg|_{\mathbf{r}=\mathbf{r}_i} \notag\\
 	 &={} & \frac{1}{16\pi(1-\nu)\mu}\frac{1}{r_{ij}} \Bigg[ \Bigg( 4(1-\nu)-\frac{4}{3} \bigg(\frac{a}{r_{ij}}\bigg)^2 \Bigg) \mathbf{\hat{r}}_{ij}\mathbf{\hat{r}}_{ij}  + \Bigg( 3-4\nu+\frac{2}{3} \bigg(\frac{a}{r_{ij}} \bigg)^2 \Bigg)(\mathbf{\underline{\hat{I}}}-\mathbf{\hat{r}}_{ij}\mathbf{\hat{r}}_{ij}) \Bigg]\cdot\mathbf{F}_j,\label{v1_1_explicit}
 \end{eqnarray} 
with $\mathbf{\hat{r}}_{ij}=(\mathbf{r}_i-\mathbf{r}_j)/r_{ij}$ the unit vector pointing from sphere $j$ to sphere $i$, see Fig.~\ref{fig1}.
Similarly, using Eqs.~(\ref{greens_function}), (\ref{u_2_0}), (\ref{w_1_1_f}), and $\nabla\times\nabla^2\mathbf{\hspace{.02cm}\underline{\hspace{-.02cm}G}}(\mathbf{r})={}\mathbf{\underline{0}}$, which follows from Eq.~(\ref{rot_u}), we find for the corresponding 
rotation of sphere $i$
\begin{equation}
	\boldsymbol{\Omega}_i^{(1)} ={} \frac{1}{2}\nabla\times\left(1+\frac{a^2}{6}\nabla^2\right)\mathbf{\hspace{.02cm}\underline{\hspace{-.02cm}G}}(\mathbf{r}-\mathbf{r}_j)\cdot\mathbf{F}_j\bigg|_{\mathbf{r}=\mathbf{r}_i}
	={} -\frac{1}{8\pi\mu r_{ij}^2}\mathbf{\hat{r}}_{ij}\times\mathbf{F}_j,\label{w1_1_explicit}
\end{equation}
see Fig.~\ref{fig1}.

To determine $\mathbf{U}_i^{(2)}$ and $\boldsymbol{\Omega}_i^{(2)}$, we first have to calculate the stresslet induced by sphere $j$ and acting onto the matrix as given by Eq.~(\ref{2s_stress}) with switched indices $i\leftrightarrow j$,
\begin{eqnarray}
	\mathbf{\underline{S}}_j^{(1)} &={} &\frac{1}{4(4-5\nu)}\frac{a^3}{r_{ij}^2}\Big[ 5(1-2\nu)(\mathbf{F}_i\mathbf{\hat{r}}_{ij} + \mathbf{\hat{r}}_{ij}\mathbf{F}_i)-3\mathbf{\underline{\hat{I}}}\,\mathbf{\hat{r}}_{ij}\cdot\mathbf{F}_i +15\mathbf{\hat{r}}_{ij}\mathbf{\hat{r}}_{ij}\mathbf{\hat{r}}_{ij}\cdot\mathbf{F}_i\Big] + \mathcal{O}(r_{ij}^{-4}).\label{2s_stress_explicit}
\end{eqnarray} 
It is sufficient to calculate $\mathbf{\underline{S}}_j^{(1)}$ to this order because $\nabla\mathbf{\hspace{.02cm}\underline{\hspace{-.02cm}G}}(\mathbf{r}-\mathbf{r}_j)$ in Eq.~(\ref{u_j_1}) is already of order $r_{ij}^{-2}$ at $\mathbf{r}=\mathbf{r}_i$.
The additional translation of sphere $i$ induced by the stresslet $\mathbf{\underline{S}}_j^{(1)}$ can now be calculated from Eqs.~(\ref{u_j_1}) and (\ref{v1_2_f}). To our desired order, we may omit the $\frac{a^2}{6}\nabla^2$-term and obtain 
\begin{equation}\label{v1_2_explicit}
	\mathbf{U}_i^{(2)} ={} -\frac{1}{32\pi(1-\nu)(4-5\nu)\mu}\frac{a^3}{r_{ij}^4}\Bigg[5(1-2\nu)^2(\mathbf{\underline{\hat{I}}} + \mathbf{\hat{r}}_{ij}\mathbf{\hat{r}}_{ij})+(37-44\nu)\mathbf{\hat{r}}_{ij}\mathbf{\hat{r}}_{ij}\Bigg]\cdot\mathbf{F}_i.
\end{equation}\end{widetext}
This expression for $\mathbf{U}_i^{(2)}$ corresponds to the lowest-order correction to the displacement of sphere $i$ resulting from a reflection of the displacement field $\mathbf{u}_i^{(0)}(\mathbf{r})$ from sphere $j$.
As for the contribution to the rotation $\boldsymbol{\Omega}_i^{(2)}$ of sphere~$i$, since $\mathbf{u}_j^{(1)}(\mathbf{r}_i)$ in Eq.~(\ref{u_j_1}) is already of order $r_{ij}^{-4}$, Eq.~(\ref{w1_2_f}) would yield an expression of higher order $\mathcal{O}(r_{ij}^{-5})$. 

As indicated above, to obtain the next-order contributions, we would have to calculate the stresslet $\mathbf{\underline{S}}_i^{(2)}$ that results from the rigidity-caused resistance of sphere $i$ in the displacement field $\mathbf{u}_j^{(1)}(\mathbf{r})$. 
This can be achieved again via Eq.~(\ref{2s_stress}) by switching the indices $(^{(0)},^{(1)})\rightarrow (^{(1)},^{(2)})$. In analogy, the resulting additional displacement field $\mathbf{u}_i^{(2)}(\mathbf{r})$ follows via Eq.~(\ref{u_i_1}) by replacing $^{(1)}\rightarrow{} ^{(2)}$, and the additional contribution $\mathbf{U}_j^{(3)}$ to the translation of sphere $j$ via Eq.~(\ref{v1_2_f}) by $(^{(1)},^{(2)},i,j)\rightarrow (^{(2)},^{(3)},j,i)$. 
Also the $\mathcal{O}(r_{ij}^{-4})$-terms in Eq.~(\ref{2s_stress_explicit}) then need to be taken into account, and the rotations $\boldsymbol{\Omega}^{(2)}_i$ contribute as well. 
This scheme can basically be continued up to an arbitrary iteration level.

Up to (including) order $r_{ij}^{-4}$, the total translation of sphere $i$ is given by $\mathbf{U}_i = \mathbf{U}_i^{(0)} + \mathbf{U}_i^{(1)} + \mathbf{U}_i^{(2)}$ and reads
\begin{widetext}
\begin{eqnarray}
	\mathbf{U}_i &={} &\Bigg\{\frac{5-6\nu}{24\pi(1-\nu)\mu a}\mathbf{\underline{\hat{I}}} -\frac{1}{32\pi(1-\nu)(4-5\nu)\mu}\frac{a^3}{r_{ij}^4}\Bigg[\Bigg(37-44\nu+10(1-2\nu)^2\Bigg)\mathbf{\hat{r}}_{ij}\mathbf{\hat{r}}_{ij}+5(1-2\nu)^2(\mathbf{\underline{\hat{I}}}- \mathbf{\hat{r}}_{ij}\mathbf{\hat{r}}_{ij}) \Bigg] \Bigg\} \cdot\mathbf{F}_i\notag\\
	&{}& + \frac{1}{16\pi(1-\nu)\mu}\frac{1}{r_{ij}} \Bigg[ \Bigg( 4(1-\nu) -\frac{4}{3} \bigg(\frac{a}{r_{ij}}\bigg)^2 \Bigg) \mathbf{\hat{r}}_{ij}\mathbf{\hat{r}}_{ij}   + \Bigg( 3-4\nu+\frac{2}{3} \bigg(\frac{a}{r_{ij}} \bigg)^2 \Bigg)(\mathbf{\underline{\hat{I}}}-\mathbf{\hat{r}}_{ij}\mathbf{\hat{r}}_{ij}) \Bigg]\cdot \mathbf{F}_j.\label{total_displacement_v_i}
\end{eqnarray} 
Similarly, the total rotation of sphere $i$ accurate up to (including) order $r_{ij}^{-4}$ is given by
\begin{equation}\label{w_rttotal}
	\boldsymbol{\Omega}_i ={} -\frac{1}{8\pi\mu r_{ij}^2}\mathbf{\hat{r}}_{ij}\times\mathbf{F}_j.
\end{equation}

So far, we have only considered two particles $i$ and $j$. However, since the governing Navier-Cauchy equations Eq.~(\ref{navier_cauchy}) are linear, we can linearly superimpose the influence of additional inclusions. That is, we simply add contributions of identical form to the right-hand sides of Eqs.~(\ref{total_displacement_v_i}) and (\ref{w_rttotal}) caused by each additional particle $j$. 

Up to (including) order $r_{ij}^{-4}$, the individual terms on the right-hand side of Eq.~(\ref{total_displacement_v_i}) then identify the components of the displaceability matrices $\mathbf{\underline{M}}_{ij}^\text{tt}$ in Eq.~(\ref{displaceability_matrix}) resulting from one- and two-body interactions \cite{puljiz2016forces} as illustrated in Figs.~\ref{fig1} and \ref{fig3}:
\begin{eqnarray}
	\mathbf{\underline{M}}_{i=j}^\text{tt} &={} & M_0^\text{t}\Bigg\{\mathbf{\underline{\hat{I}}} -  \sum_{\scriptsize\begin{aligned}k \! &=\! 1 \\[-4pt] k\! &\ne \! i\end{aligned}}^N 
	 \frac{3}{4(4-5\nu)(5-6\nu)}\bigg(\frac{a}{r_{ik}}\bigg)^4\Bigg[ \Bigg(37-44\nu+10(1-2\nu)^2\Bigg)\mathbf{\hat{r}}_{ik}\mathbf{\hat{r}}_{ik} +5(1-2\nu)^2(\mathbf{\underline{\hat{I}}}- \mathbf{\hat{r}}_{ik}\mathbf{\hat{r}}_{ik}) \Bigg] \Bigg\},\quad \label{M_II_2}\\
	\mathbf{\underline{M}}_{i\not=j}^\text{tt} &={} & M_0^\text{t}  \frac{3}{2(5-6\nu)}\frac{a}{r_{ij}}  \Bigg[ \Bigg( 4(1-\nu)-\frac{4}{3} \bigg(\frac{a}{r_{ij}}\bigg)^2 \Bigg) \mathbf{\hat{r}}_{ij}\mathbf{\hat{r}}_{ij}   + \Bigg( 3-4\nu+\frac{2}{3} \bigg(\frac{a}{r_{ij}} \bigg)^2 \Bigg)(\mathbf{\underline{\hat{I}}}-\mathbf{\hat{r}}_{ij}\mathbf{\hat{r}}_{ij}) \Bigg]    +    \mathbf{\underline{M}}^{\text{tt}(3)}_{i\not=j}
\label{M_IJ_2}    ,
\end{eqnarray}\end{widetext}
with $i,j\in\{1,2,...,N\}$ and
\begin{equation}
	M_0^\text{t} ={} \frac{5-6\nu}{24\pi(1-\nu)\mu a}.\label{M_0_t}
\end{equation}
The contribution $\mathbf{\underline{M}}^{\text{tt}(3)}_{i\not=j}$ represents three-body interactions and will be separately derived in Sec.~\ref{Section_9_three_spheres}.

Furthermore, from Eq.~(\ref{w_rttotal}) we find for the components of the rotateability matrices $\mathbf{\underline{M}}_{ij}^\text{rt}$ up to (including) order $r_{ij}^{-4}$
\begin{eqnarray}
	\mathbf{\underline{M}}_{i=j}^\text{rt} &={} & \mathbf{\underline{0}},\label{M_II_rt}\\
	\mathbf{\underline{M}}_{i\not=j}^\text{rt} &={} & -M_0^\text{r} \frac{\mathbf{\hat{r}}_{ij}}{r_{ij}^2}\times,\label{M_IJ_rt}
\end{eqnarray}
see Fig.~\ref{fig1}, with
\begin{equation}\label{M_0_r}
	M_0^\text{r} ={} \frac{1}{8\pi\mu}.
\end{equation}

\subsection{Torques externally imposed on or induced between the inclusions}

\begin{figure}
\centerline{\includegraphics[width=\columnwidth]{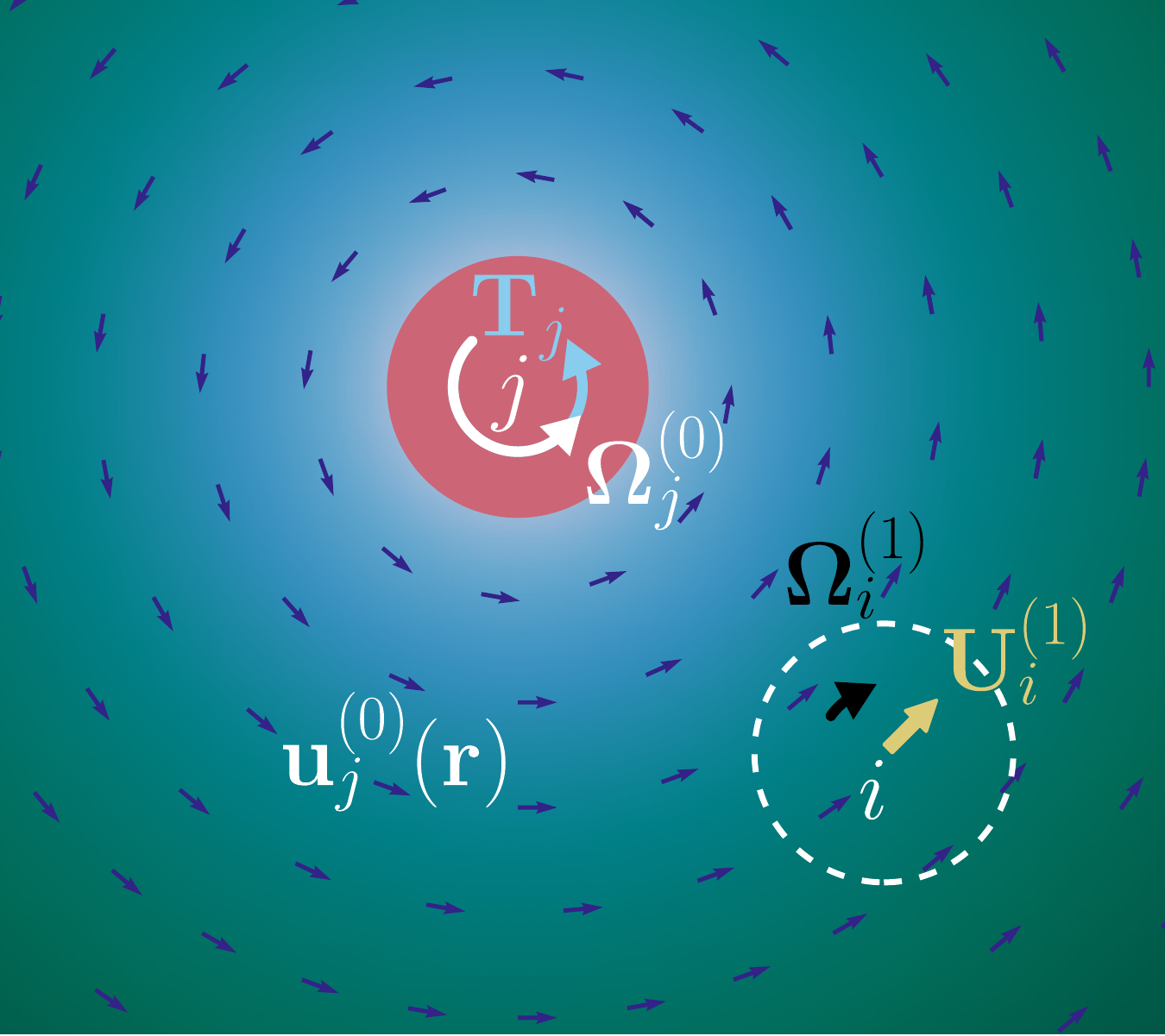}}
\caption{
Illustration of the immediate effect that the rotation of sphere $j$ has on the translation and rotation of another sphere $i$.
A torque $\mathbf{T}_j$ is externally imposed onto sphere $j$ that, as a consequence, gets rigidly rotated by $\boldsymbol{\Omega}_j^{(0)}$, see Eq.~(\ref{omega_2_spheres2}).
Moreover, the surrounding matrix is distorted, as described by the displacement field $\mathbf{u}_j^{(0)}(\mathbf{r})$, see Eq.~(\ref{u_j_0_rotated_spheres}).
The local directions of $\mathbf{u}_j^{(0)}(\mathbf{r})$ are marked by the small normalized arrows. We indicated the local magnitude of $\mathbf{u}_j^{(0)}(\mathbf{r})$ by the background color, where brighter color represents higher magnitude and the color values follow an arc-tangent scale.
Sphere $i$ is exposed to the induced displacement field $\mathbf{u}_j^{(0)}(\mathbf{r})$ and therefore gets translated by $\mathbf{U}_i^{(1)}$ and rotated by $\boldsymbol{\Omega}_i^{(1)}$, see Eqs.~(\ref{v_i_1_rotated}) and (\ref{w_i_1_rotated}), respectively.
Explicit results are given in Eqs.~(\ref{v_i_1_rotated_expl}) and (\ref{w_i_1_rotated_expl}).
Overall, in this way we obtain the corresponding contributions to the \textit{displaceability and rotateability matrices} $\mathbf{\underline{M}}_{i=j}^\text{tr}$, $\mathbf{\underline{M}}_{i\neq j}^\text{tr}$, $\mathbf{\underline{M}}_{i=j}^\text{rr}$, and $\mathbf{\underline{M}}_{i\neq j}^\text{rr}$ in Eqs.~(\ref{M_II_tr})--(\ref{M_IJ_rr}), respectively, up to inverse quartic order in the particle distances.
}
\label{fig4}
\end{figure}

Instead of forces $\mathbf{F}_i$ and $\mathbf{F}_j$, let us now consider torques $\mathbf{T}_i$ and $\mathbf{T}_j$ externally imposed on or induced between two rigid spherical inclusions $i$ and $j$. The treatment of this situation \mpu{follows the same lines}, therefore we will be significantly briefer here. 

To zeroth order, where matrix-mediated interactions between the two spheres are ignored, the torques cause rotations $\boldsymbol{\Omega}_i^{(0)}$ and $\boldsymbol{\Omega}_j^{(0)}$ of the particles, respectively, which follow via Eq.~(\ref{torque_rotation}) as
\begin{eqnarray}
	\boldsymbol{\Omega}_i^{(0)} &=&{} \frac{1}{8\pi\mu a^3} \mathbf{T}_i, 
\label{omega_2_spheres1}
\\
 	\boldsymbol{\Omega}_j^{(0)} &=&{} \frac{1}{8\pi\mu a^3} \mathbf{T}_j.
\label{omega_2_spheres2}
\end{eqnarray}
Due to the stick boundary conditions, the rotated spheres drag the surrounding matrix along and therefore generate displacement fields as given by Eq.~(\ref{displacement_rotated_sphere}),
\begin{eqnarray}
	\mathbf{u}_i^{(0)}(\mathbf{r}) &={} &\left(\frac{a}{|\mathbf{r}-\mathbf{r}_i|}\right)^{\!3}\;\boldsymbol{\Omega}_i^{(0)}\times(\mathbf{r}-\mathbf{r}_i), \label{u_i_0_rotated_spheres}\\
	\mathbf{u}_j^{(0)}(\mathbf{r}) &={} &\left(\frac{a}{|\mathbf{r}-\mathbf{r}_j|}\right)^{\!3}\;\boldsymbol{\Omega}_j^{(0)}\times(\mathbf{r}-\mathbf{r}_j),\label{u_j_0_rotated_spheres}
\end{eqnarray}
see Fig.~\ref{fig4}.

Similarly to the case of translated spheres, the displacement field $\mathbf{u}_j^{(0)}(\mathbf{r})$ resulting from the rotation of sphere $j$ affects the total displacement and rotation of sphere $i$. 
Moreover, due to its rigidity, additional stresses occur when sphere $i$ resists deformations that would be induced by the displacement field $\mathbf{u}_j^{(0)}(\mathbf{r})$. 
The induced translation $\mathbf{U}_i^{(1)}$, additional rotation $\boldsymbol{\Omega}_i^{(1)}$, and rigidity-based stresslet $\mathbf{\underline{S}}_i^{(1)}$ exerted by sphere $i$ can be calculated using Eqs.~(\ref{faxen_f}), (\ref{faxen_r}), and (\ref{faxen_3}), respectively. There, $\mathbf{u}_j^{(0)}(\mathbf{r})$ is inserted as the imposed displacement field. 
We find
\begin{widetext}
\begin{eqnarray}
	\mathbf{U}_i^{(1)} &={} &\bigg(1+\frac{a^2}{6}\nabla^2\bigg)\mathbf{u}_j^{(0)}(\mathbf{r})\bigg|_{\mathbf{r}=\mathbf{r}_i}, \label{v_i_1_rotated}\\
	\boldsymbol{\Omega}_i^{(1)} &={} &\frac{1}{2}\nabla\times\mathbf{u}_j^{(0)}(\mathbf{r})\bigg|_{\mathbf{r}=\mathbf{r}_i}, \label{w_i_1_rotated}\\
	\mathbf{\underline{S}}_i^{(1)} &={} &-\frac{4\pi(1-\nu)\mu a^3}{4-5\nu}\left(1+\frac{a^2}{10}\nabla^2\right)\Bigg[\frac{1}{1-2\nu}\mathbf{\underline{\hat{I}}}\nabla\cdot\mathbf{u}_j^{(0)}(\mathbf{r})+ \frac{5}{2}\Big(\nabla\mathbf{u}_j^{(0)}(\mathbf{r})+\big(\nabla\mathbf{u}_j^{(0)}(\mathbf{r})\big)^T\Big) \Bigg]\Bigg|_{\mathbf{r}=\mathbf{r}_i}.
\end{eqnarray}\end{widetext}
Analogously to Eq.~(\ref{u_i_1}), the displacement field resulting from the rigidity-based resistance of sphere~$i$ against deformation is given by
\begin{equation}
	\mathbf{u}_i^{(1)}(\mathbf{r}) ={} -(\mathbf{\underline{S}}_i^{(1)}\cdot\nabla)\cdot\mathbf{\hspace{.02cm}\underline{\hspace{-.02cm}G}}(\mathbf{r}-\mathbf{r}_i).
\end{equation}
Since the stresslet $\mathbf{\underline{S}}_i^{(1)}$ here yields an expression of order $r_{ij}^{-3}$, $\mathbf{u}_i^{(1)}(\mathbf{r}_j)$ is already of order $r_{ij}^{-5}$. Therefore, we can stop our \mpu{iteration at this point}, confining ourselves to contributions up to (including) order $r_{ij}^{-4}$. 
Again, all corresponding expressions for sphere $j$ are obtained by simply switching all indices $i\leftrightarrow j$.

To derive explicit analytical expressions, we insert Eqs.~(\ref{omega_2_spheres2}) and (\ref{u_j_0_rotated_spheres}) into Eqs.~(\ref{v_i_1_rotated}) and (\ref{w_i_1_rotated}). We obtain
\begin{eqnarray}
	\mathbf{U}_i^{(1)} 	&={} & -\frac{1}{8\pi\mu r_{ij}^2}\mathbf{\hat{r}}_{ij}\times\mathbf{T}_j,\label{v_i_1_rotated_expl}\\
	\boldsymbol{\Omega}_i^{(1)} &={} & \frac{1}{16\pi\mu r_{ij}^3}\big[3\mathbf{\hat{r}}_{ij}\mathbf{\hat{r}}_{ij}-\mathbf{\underline{\hat{I}}}\big]\cdot\mathbf{T}_j,\label{w_i_1_rotated_expl}
\end{eqnarray}
as illustrated in Fig.~\ref{fig4}.
From Eq.~(\ref{v_i_1_rotated_expl}), 
we see that an additional translation of sphere $i$ only occurs, if $\mathbf{\hat{r}}_{ij}$ is not (anti)parallel to $\mathbf{T}_j$. 
Moreover, sphere $i$ is translated in the same direction as the nearest surface point of sphere $j$. 
The sense of the additional rotation $\boldsymbol{\Omega}_i^{(1)}$ that only vanishes at infinite particle separation $r_{ij}$ depends on the relative angular configuration according to Eq.~(\ref{w_i_1_rotated_expl}). 
For instance, if $\mathbf{\hat{r}}_{ij}\parallel\mathbf{T}_j$, i.e., both spheres and the imposed torque $\mathbf{T}_j$ align along a common axis, then the zero-order rotation $\boldsymbol{\Omega}_j^{(0)}$ and the additional rotation $\boldsymbol{\Omega}_i^{(1)}$ have the same sense. For $\mathbf{\hat{r}}_{ij}\perp\mathbf{T}_j$, i.e., the imposed torque $\mathbf{T}_j$ is perpendicular to the plane that contains both spheres, these two rotations have opposite sense.

Overall, the total translation of sphere $i$ to our desired order is given by $\mathbf{U}_i^{(1)}$ in Eq.~(\ref{v_i_1_rotated_expl}). The total rotation up to (including) order $r_{ij}^{-4}$ equals $\boldsymbol{\Omega}_i^{(0)}+\boldsymbol{\Omega}_i^{(1)}$, see Eqs.~(\ref{omega_2_spheres1}) and (\ref{w_i_1_rotated_expl}). 
Therefore, with the same reasoning as in Sec.~\ref{subsection_forces_imposed}, we can read off the components of the corresponding displaceability matrices $\mathbf{\underline{M}}_{ij}^\text{tr}$ and rotateability matrices $\mathbf{\underline{M}}_{ij}^\text{rr}$ from Eqs.~(\ref{omega_2_spheres1}), (\ref{v_i_1_rotated_expl}), and (\ref{w_i_1_rotated_expl}) as
\begin{eqnarray}
	\mathbf{\underline{M}}_{i=j}^\text{tr} &={} &\mathbf{\underline{0}},\label{M_II_tr}\\
	\mathbf{\underline{M}}_{i\not=j}^\text{tr} &={} &-M_0^\text{r}\frac{\mathbf{\hat{r}}_{ij}}{r_{ij}^2}\times,\label{M_IJ_tr}\\
	\mathbf{\underline{M}}_{i=j}^\text{rr} &={} &M_0^\text{r}\frac{1}{a^3}\mathbf{\underline{\hat{I}}},\label{M_II_rr}\\
	\mathbf{\underline{M}}_{i\not=j}^\text{rr} &={} &M_0^\text{r}\frac{1}{2r_{ij}^3}\big[3\mathbf{\hat{r}}_{ij}\mathbf{\hat{r}}_{ij}-\mathbf{\underline{\hat{I}}}\big],\label{M_IJ_rr}
\end{eqnarray}
where $M_0^\text{r}$ was introduced in Eq.~(\ref{M_0_r}). See also the illustration in Fig.~\ref{fig4}. Based on the linearity of the governing Navier-Cauchy equations in Eq.~(\ref{navier_cauchy}), we may sum up the influence of imposed or induced forces in Sec.~\ref{subsection_forces_imposed} and the ones just derived for imposed or induced torques and combine them in an overall matrix equation as given in Eq.~(\ref{displaceability_matrix}).

\section{Three-body interactions}\label{Section_9_three_spheres}

Following the same strategy as in Sec.~\ref{Section_8_two_rigid_sphere_interaction}, we now derive similar expressions for the three-body interactions. In this way, we determine the components of the matrix $\mathbf{\underline{M}}^{\text{tt}(3)}_{i\not=j}$ in Eq.~(\ref{M_IJ_2}). Again, we adapt the procedure for low-Reynolds-number hydrodynamics presented in Ref.~\onlinecite{dhont1996introduction}. 

For this purpose, we now consider three rigid spherical inclusions of radius $a$, located at positions $\mathbf{r}_i$, $\mathbf{r}_j$, and $\mathbf{r}_k$. They are acted on by externally imposed or induced forces $\mathbf{F}_i$, $\mathbf{F}_j$, and $\mathbf{F}_k$, respectively.
To zeroth order, i.e., not taking into account matrix-mediated interactions between the inclusions, sphere $i$ creates a displacement field as given by Eq.~(\ref{u_1_0}). Corresponding expressions follow for spheres $j$ and $k$ by switching indices $i\rightarrow j$ and $i\rightarrow k$, respectively.  

In analogy to Eq.~(\ref{v_1_1_f}), we can calculate from the first Fax\'en law Eq.~(\ref{faxen_f}) the translation that sphere $i$ acquires within the linearly superimposed displacement fields $\mathbf{u}_j^{(0)}(\mathbf{r})$ and $\mathbf{u}_k^{(0)}(\mathbf{r})$. Using $\mathbf{u}_j^{(0)}(\mathbf{r})+\mathbf{u}_k^{(0)}(\mathbf{r})$ as the imposed field on the right-hand side of Eq.~(\ref{faxen_f}), we obtain
\begin{equation}
	\mathbf{U}_i^{(1)} ={} \left(1+\frac{a^2}{6}\nabla^2\right)\bigg[\mathbf{u}_j^{(0)}(\mathbf{r}) + \mathbf{u}_k^{(0)}(\mathbf{r})\bigg]\bigg|_{\mathbf{r}=\mathbf{r}_i}. \label{v_i_1_3}
\end{equation}
\mpu{Corresponding} expressions follow for spheres $j$ and $k$ by switching in this equation $i\leftrightarrow j$ and $i\leftrightarrow k$, respectively. 

Again, sphere $i$ resists any deformation that would be implied by the matrix deformations described by $\mathbf{u}_j^{(0)}(\mathbf{r})$ and $\mathbf{u}_k^{(0)}(\mathbf{r})$. 
The resulting stresslet that sphere $i$ thus exerts onto the matrix can be calculated in analogy to Eq.~(\ref{2s_stress}) and using Eq.~(\ref{faxen_3}), 
\begin{widetext}
\begin{eqnarray}
	\mathbf{\underline{S}}_i^{(1)} &={} &-\frac{4\pi(1-\nu)\mu a^3}{4-5\nu}\left(1+\frac{a^2}{10}\nabla^2\right)\left[\frac{1}{1-2\nu}\mathbf{\underline{\hat{I}}}\nabla\cdot\big (\mathbf{u}_j^{(0)}(\mathbf{r})+ \mathbf{u}_k^{(0)}(\mathbf{r})\big)\right.\notag\\
	&{} &+ \frac{5}{2}\bigg(\nabla\big(\mathbf{u}_j^{(0)}(\mathbf{r}) + \mathbf{u}_k^{(0)}(\mathbf{r})\big) + \Big(\nabla\big(\mathbf{u}_j^{(0)}(\mathbf{r})+ \mathbf{u}_k^{(0)}(\mathbf{r})\big)\Big)^T\bigg) \bigg]\bigg|_{\mathbf{r}=\mathbf{r}_i}.
\label{threebody_S_i_1}
\end{eqnarray}\end{widetext}
It produces the displacement field
\begin{equation}
	\mathbf{u}_i^{(1)}(\mathbf{r}) ={} -(\mathbf{\underline{S}}_i^{(1)}\cdot\nabla)\cdot\mathbf{\hspace{.02cm}\underline{\hspace{-.02cm}G}}(\mathbf{r}-\mathbf{r}_i),\label{u_i_1_3}
\end{equation}
see Eq.~(\ref{multipole_fts}), due to the resistance of sphere $i$ to deformations implied by $\mathbf{u}_j^{(0)}(\mathbf{r})$ and $\mathbf{u}_k^{(0)}(\mathbf{r})$. Once more, expressions for spheres $j$ and $k$ are obtained from this equation by replacing $i\rightarrow j$ and $i\rightarrow k$, respectively.

Next, we use the sum of the resulting displacement fields $\mathbf{u}_j^{(1)}(\mathbf{r})+\mathbf{u}_k^{(1)}(\mathbf{r})$ as the imposed field on the right-hand side of Fax\'en's first law, Eq.~(\ref{faxen_f}). In this way, we can calculate the
additional translation $\mathbf{U}_i^{(2)}$ that sphere $i$ experiences in these rigidity-induced displacement fields,
\begin{equation}
	\mathbf{U}_i^{(2)} ={} \left(1+\frac{a^2}{6}\nabla^2\right)\Big[\mathbf{u}_j^{(1)}(\mathbf{r}) + \mathbf{u}_k^{(1)}(\mathbf{r})\Big]\bigg|_{\mathbf{r}=\mathbf{r}_i}. \label{v_i_2_3}
\end{equation}

At first glance, this expression is of identical shape as Eq.~(\ref{v1_2_f}) for the two-sphere interaction. The only difference seems to be that here we take into account the two contributions from the two spheres $j$ and $k$, instead of only one.
Indeed, we here recover all contributions that we have already identified in Sec.~\ref{subsection_forces_imposed}.
However, there is now more to that.

For simplicity, let us for the moment only consider in Eq.~(\ref{v_i_2_3}) the effect of the displacement field $\mathbf{u}_k^{(1)}(\mathbf{r})$, where the latter according to Eq.~(\ref{u_i_1_3}) is given by
\begin{equation}
	\mathbf{u}_k^{(1)}(\mathbf{r}) ={} -(\mathbf{\underline{S}}_k^{(1)}\cdot\nabla)\cdot\mathbf{\hspace{.02cm}\underline{\hspace{-.02cm}G}}(\mathbf{r}-\mathbf{r}_k).
\label{u1_fuer_k}
\end{equation}
Here, $\mathbf{\underline{S}}_k^{(1)}$ is the stresslet that sphere $k$ exerts onto the surrounding matrix 
due to its rigidity. 
It arises as sphere $k$ opposes to deformations implied by $\mathbf{u}_i^{(0)}(\mathbf{r})$ and $\mathbf{u}_j^{(0)}(\mathbf{r})$. The latter displacement fields directly result from the external forces $\mathbf{F}_i$ and $\mathbf{F}_j$ acting onto spheres $i$ and $j$, respectively. 
These two forces lead to two different scenarios. 

The first scenario has already been described in Sec.~\ref{subsection_forces_imposed}. 
A force $\mathbf{F}_i$ acting onto sphere $i$ generates the displacement field $\mathbf{u}_i^{(0)}(\mathbf{r})$. This field is \textit{reflected} by sphere $k$. Then it acts back onto sphere $i$ in the form of $\mathbf{u}_k^{(1)}(\mathbf{r})$, contributing to $\mathbf{U}^{(2)}_i$ in Eq.~(\ref{v_i_2_3}). We abbreviate this chain of matrix-mediated interactions as $i\leftarrow k\leftarrow i$. 

In the second scenario, a force $\mathbf{F}_j$ acting onto a \textit{third} sphere $j$ induces a displacement field $\mathbf{u}_j^{(0)}(\mathbf{r})$. 
This field is then reflected by sphere $k$ due to its rigidity in the form of $\mathbf{u}_k^{(1)}(\mathbf{r})$. 
However, in the present three-body configuration, the reflected field also affects sphere $i$ and contributes to its displacement $\mathbf{U}_i^{(2)}$ in Eq.~(\ref{v_i_2_3}). 
This \textit{three-body interaction} 
thus defines a further contribution 
in addition to the pairwise interactions derived in Sec.~\ref{subsection_forces_imposed}. 
We abbreviate the corresponding chain of matrix-mediated interactions as $i\leftarrow k\leftarrow j$.

Altogether, we find two such three-body interactions contributing to $\mathbf{U}_i^{(2)}$ in Eq.~(\ref{v_i_2_3}) in addition to the pairwise interactions. 
The first one works as described, $i\leftarrow k\leftarrow j$, and we denote it as $\mathbf{U}_{ikj}^{(2)}$.
The second one works via $i\leftarrow j \leftarrow k$, which would then be termed $\mathbf{U}_{ijk}^{(2)}$.
Explicit calculation yields
\begin{widetext}
\begin{eqnarray}
\mathbf{U}_{ikj}^{(2)} &={} &-\left(1+\frac{a^2}{6}\nabla^2\right)(\mathbf{\underline{S}}_k^{(1)}\cdot\nabla)\cdot\mathbf{\hspace{.02cm}\underline{\hspace{-.02cm}G}}(\mathbf{r}-\mathbf{r}_k)\bigg|_{\mathbf{r}=\mathbf{r}_i}\notag\\
	&={} &\frac{1}{64\pi(1-\nu)(4-5\nu)\mu}\frac{a^3}{r_{ik}^2r_{jk}^2} \bigg[
	-10(1-2\nu)\bigg( (1-2\nu)\Big((\mathbf{\hat{r}}_{ik}\cdot\mathbf{\hat{r}}_{jk}) \mathbf{\underline{\hat{I}}} + \mathbf{\hat{r}}_{jk}\mathbf{\hat{r}}_{ik} \Big) \notag\\
	&{}& + 3 (\mathbf{\hat{r}}_{ik}\cdot\mathbf{\hat{r}}_{jk}) (\mathbf{\hat{r}}_{ik}\mathbf{\hat{r}}_{ik} + \mathbf{\hat{r}}_{jk}\mathbf{\hat{r}}_{jk}) - \mathbf{\hat{r}}_{ik}\mathbf{\hat{r}}_{jk} \bigg) + 3\Big(7-4\nu - 15 (\mathbf{\hat{r}}_{ik}\cdot\mathbf{\hat{r}}_{jk})^2\Big)\mathbf{\hat{r}}_{ik}\mathbf{\hat{r}}_{jk}
	\bigg]\cdot\mathbf{F}_j + \mathcal{O}\left((r_{ik},r_{jk})^{-5}\right).\quad \label{v_3b}
\end{eqnarray} 
$\mathbf{U}_{ijk}^{(2)}$ is readily obtained from this expression by switching indices $j\leftrightarrow k$.

In summary, to our desired order, i.e., up to (including) quartic order in the inverse particle separation distances, two- and three-body interactions contribute to $\mathbf{U}_i^{(2)}$. 
\mpu{The latter} follow from Eq.~(\ref{v_3b}) for $i\neq j$.
For $i=j$, Eq.~(\ref{v_3b}) exactly reproduces the two-body contributions listed already in Eq.~(\ref{v1_2_explicit}).
Again due to the linearity of the governing elasticity equations Eq.~(\ref{navier_cauchy}), we may simply add the additional contributions $\mathbf{U}_{ijk}^{(2)}$ and $\mathbf{U}_{ikj}^{(2)}$ to our previous explicit analytical expression for the overall displacement of sphere $i$. 

Superimposing all contributions that result for the coupled displacements and rotations of $N$ identical spherical inclusions, we return to our formalism in terms of the displaceability and rotateability matrices in Eq.~(\ref{displaceability_matrix}). We can now read off from Eq.~(\ref{v_3b}) the additional \textit{three-body contribution} $\mathbf{\underline{M}}^{\text{tt}(3)}_{i\not=j}$ to the \textit{displaceability matrix} in Eq.~(\ref{M_IJ_2}) \cite{puljiz2016forces}, 
\begin{eqnarray}\label{M_tt_3}
	\mathbf{\underline{M}}_{i\not=j}^\text{tt(3)} &={} & M_0^\text{t}\frac{3}{8(4-5\nu)(5-6\nu)}
	\sum_{\scriptsize\begin{aligned}k\! &=\! 1 \\[-4pt] k\! &\ne \! i,\!j\end{aligned}}^N
	\bigg(\frac{a}{r_{ik}}\bigg)^2 
	\bigg(\frac{a}{r_{jk}}\bigg)^2
	\bigg[
		-10(1-2\nu)\Big( (1-2\nu)((\mathbf{\hat{r}}_{ik}\cdot\mathbf{\hat{r}}_{jk}) \mathbf{\underline{\hat{I}}}+ \mathbf{\hat{r}}_{jk}\mathbf{\hat{r}}_{ik} ) \notag\\
				&{} &
		 + 3 (\mathbf{\hat{r}}_{ik}\cdot\mathbf{\hat{r}}_{jk}) (\mathbf{\hat{r}}_{ik}\mathbf{\hat{r}}_{ik} + \mathbf{\hat{r}}_{jk}\mathbf{\hat{r}}_{jk}) - \mathbf{\hat{r}}_{ik}\mathbf{\hat{r}}_{jk} \Big) + 3\Big(7-4\nu - 15 (\mathbf{\hat{r}}_{ik}\cdot\mathbf{\hat{r}}_{jk})^2\Big)\mathbf{\hat{r}}_{ik}\mathbf{\hat{r}}_{jk}
		\bigg]
\end{eqnarray}\end{widetext}
where $M_0^\text{t}$ was introduced in Eq.~(\ref{M_0_t}). This expression is exact 
up to (including) order $(r_{ik}, r_{jk})^{-4}$. 

It can readily be seen that rotations caused by three-body interactions are of higher order than $(r_{ik}, r_{jk})^{-4}$. 
The additional rotation $\boldsymbol{\Omega}_i^{(2)}$ of sphere $i$ due to 
the reflected displacement fields $\mathbf{u}_j^{(1)}(\mathbf{r})$ and $\mathbf{u}_k^{(1)}(\mathbf{r})$ follows from Fax\'en's second law Eq.~(\ref{faxen_r}) and reads
\begin{equation}
	\boldsymbol{\Omega}_i^{(2)} ={} \frac{1}{2}\nabla\times\Big[\mathbf{u}_j^{(1)}(\mathbf{r})+\mathbf{u}_k^{(1)}(\mathbf{r})\Big]\bigg|_{\mathbf{r}=\mathbf{r}_i}.\label{wi_2_3_f}
\end{equation}
This expression is already of order $(r_{ik},r_{jk})^{-5}$, because both $\mathbf{u}_j^{(1)}(\mathbf{r}_i)$ and $\mathbf{u}_k^{(1)}(\mathbf{r}_i)$ are of order $(r_{ik},r_{jk})^{-4}$, which is obtained by combining Eqs.~(\ref{greens_function}), (\ref{u_1_0}), (\ref{threebody_S_i_1}), (\ref{u_i_1_3}), and (\ref{u1_fuer_k}). 
Therefore, to our desired order, we find
\begin{equation}
    \mathbf{\underline{M}}_{i\not=j}^\text{rt(3)} = \mathbf{\underline{0}}.
\label{M_rt_3}
\end{equation}

Similarly, we do not obtain any three-body contribution to 
the remaining displaceability and rotateability matrices up to our desired order. Reconsidering the above derivation of $\mathbf{\underline{M}}_{i\not=j}^\text{tt(3)}$, we find that 
solely the lowest-order parts of all contributing expressions finally enter Eq.~(\ref{M_tt_3}). When torques are externally imposed on or induced between the individual spheres, already the resulting zero-order displacement fields are one order higher in the inverse separation distances. This follows by comparing Eqs.~(\ref{omega_2_spheres1})--(\ref{u_j_0_rotated_spheres}) to the case of imposed or induced forces, see Eqs.~(\ref{greens_function}), (\ref{u_1_0}), and (\ref{u_2_0}). Therefore, the reflected displacement fields due to the rigidity of the spherical inclusions, see Eqs.~(\ref{threebody_S_i_1}) and (\ref{u_i_1_3}), already yield expressions of order $(r_{ik},r_{jk})^{-5}$. Thus we find to our desired order
\begin{eqnarray}
	\mathbf{\underline{M}}_{i\not=j}^\text{tr(3)} &={} & 
        \mathbf{\underline{0}},
\label{M_tr_3}\\
	\mathbf{\underline{M}}_{i\not=j}^\text{rr(3)} &={} & 
        \mathbf{\underline{0}}.
\label{M_rr_3}
\end{eqnarray}

These results complete our derivation of the displaceability and rotateability matrices up to (including) inverse quartic order in the separation distances between the individual spherical particles.

\mpu{Naturally, for larger deformations, the more the nonlinearities in the elastic response of the embedding matrix become significant, the less exact our approach will become. In nonlinearly elastic situations, if an exact quantitative evaluation is necessary, simulations are still mandatory. Yet, for a first and quick qualitative scan in the absence of bifurcational behavior, our analytical expressions will in many cases be helpful. Moreover, our approach may still be valuable to significantly speed up corresponding simulations. For this purpose, the configuration calculated from our linearly elastic formalism could be used as an initialization in iterative simulation methods.}

\section{Some illustrative examples}\label{Section_10_rotations}

\begin{figure}
\centerline{\includegraphics[width=\columnwidth]{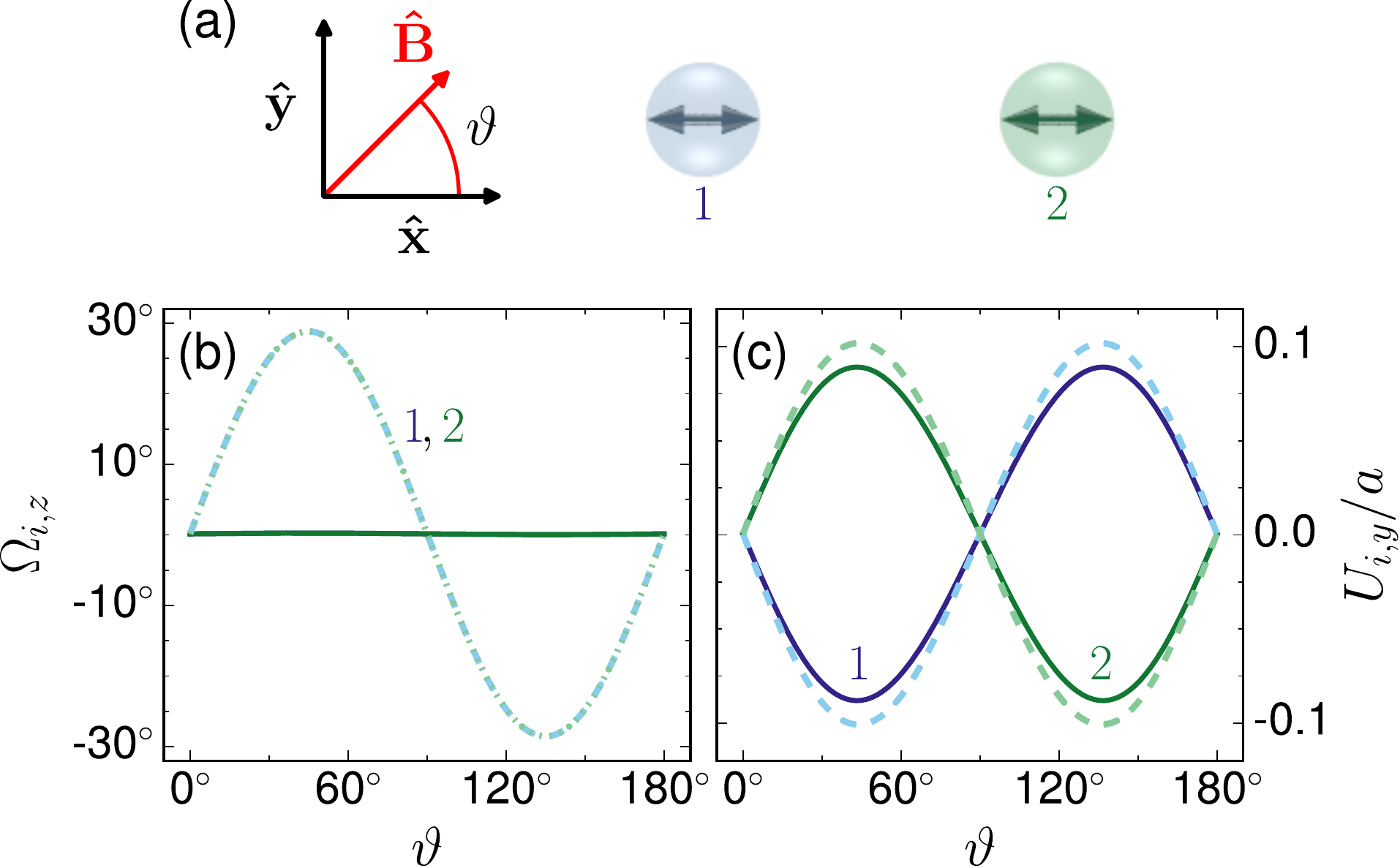}}
\caption{
\mpu{(a) Schematic illustration of the modified initial spatial configuration of a two-particle system that had been investigated before in Ref.~\onlinecite{puljiz2016forces} in the absence of induced torques. The double arrows indicate the initial orientations of additional magnetic anisotropy axes.
$\vartheta$ here is defined as the angle between the unit vector $\mathbf{\hat{x}}$ and the direction $\mathbf{\hat{B}}$ of an external magnetic field (right-handed system). This external magnetic field is initially applied parallel to $\mathbf{\hat{x}}$ and then rotated counterclockwise in the $xy$-plane until $\vartheta=180^\circ$. Magnetic forces arise between the particles as given by Eq.~(\ref{magnetic_force}) due to induced magnetic moments $\mathbf{m}\parallel\mathbf{B}$.
(b) Plot of the $z$-components of the rotation vectors $\boldsymbol{\Omega}_i$ of the particles as functions of $\vartheta$. In this configuration, all rotations occur in the $xy$-plane, therefore all other components of $\boldsymbol{\Omega}_i$ vanish.
The continuous line represents the rotations of particles $1$ and $2$, if induced torques are set to zero. The dashed and dotted lines show the results when the torques are included as they result from Eqs.~(\ref{eqSW}) and (\ref{stoner-wohlfarth_torque}).
The maximum magnitudes of rotation occur around $\vartheta=45^\circ$ and $135^\circ$, respectively, as expected from the underlying Stoner-Wohlfarth model, with opposite signs beyond $90^\circ$ because the anisotropy axes do not have any preferred direction.
(c) The $y$-components of the displacement vectors $\mathbf{U}_i$ without (continuous lines) and with (dashed) inclusion of the torques. The curves are labeled by the particle numbers, see (a).
In this set-up, the torques amplify the magnitudes of the displacements due to their sense of rotation.}
}
\label{fig5}
\end{figure}

\begin{figure}
\centerline{\includegraphics[width=\columnwidth]{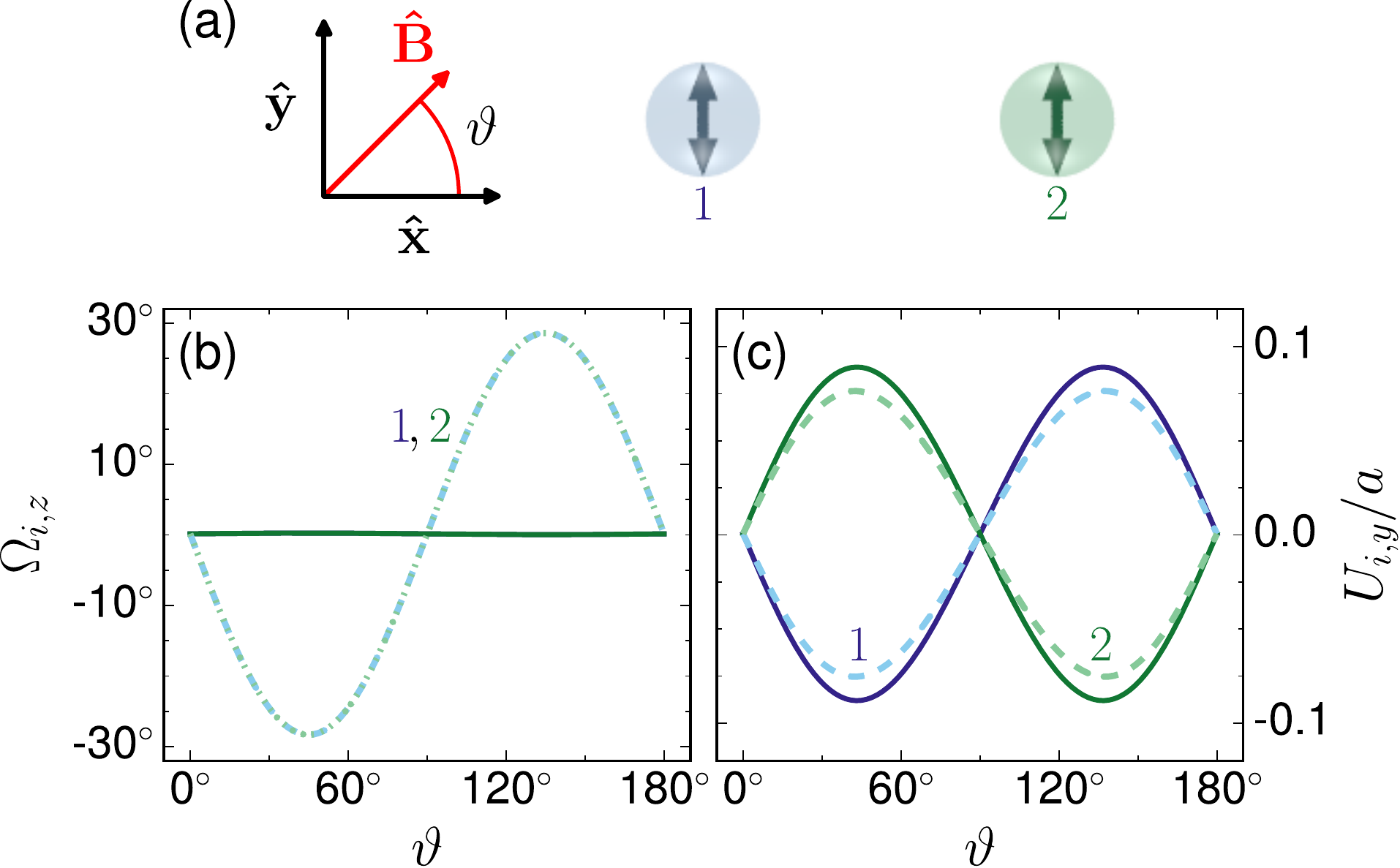}}
\caption{
The same as in Fig.~\ref{fig5}, except for the orientations of the anisotropy axes. They are now initially oriented along $\mathbf{\hat{y}}$, see (a). 
Therefore, the particles are rotated inversely when compared to Fig.~\ref{fig5}, as shown by the dashed and dotted lines in (b).
Overall, this now leads to an attenuation of the displacements in $\mathbf{\hat{y}}$-direction (c).
}
\label{fig6}
\end{figure}

\begin{figure}
\centerline{\includegraphics[width=\columnwidth]{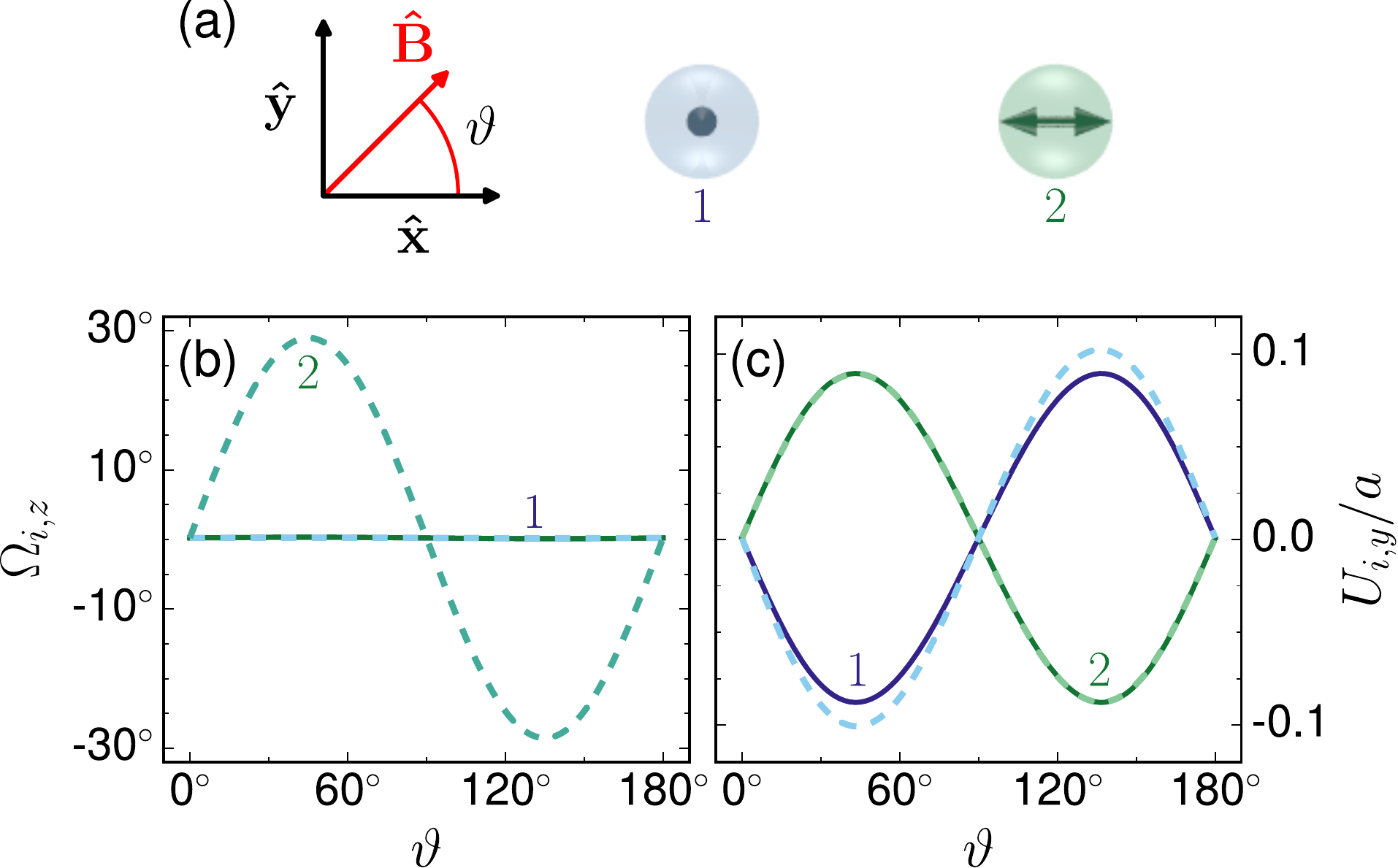}}
\caption{
The same as in Fig.~\ref{fig5}, but now the anisotropy axis of particle $1$ is along $\mathbf{\hat{z}}$, see (a). 
(b) Then, the induced torque $\mathbf{T}_1$ vanishes for all $\vartheta$ and particle $1$ is only weakly rotated due to the rotation--translation coupling in Eq.~(\ref{M_IJ_rt}).
Therefore, we do not observe a change in the displacements $U_{2,y}$ in (c) when the torques are included.
}
\label{fig7}
\end{figure}t

\begin{figure}
\centerline{\includegraphics[width=\columnwidth]{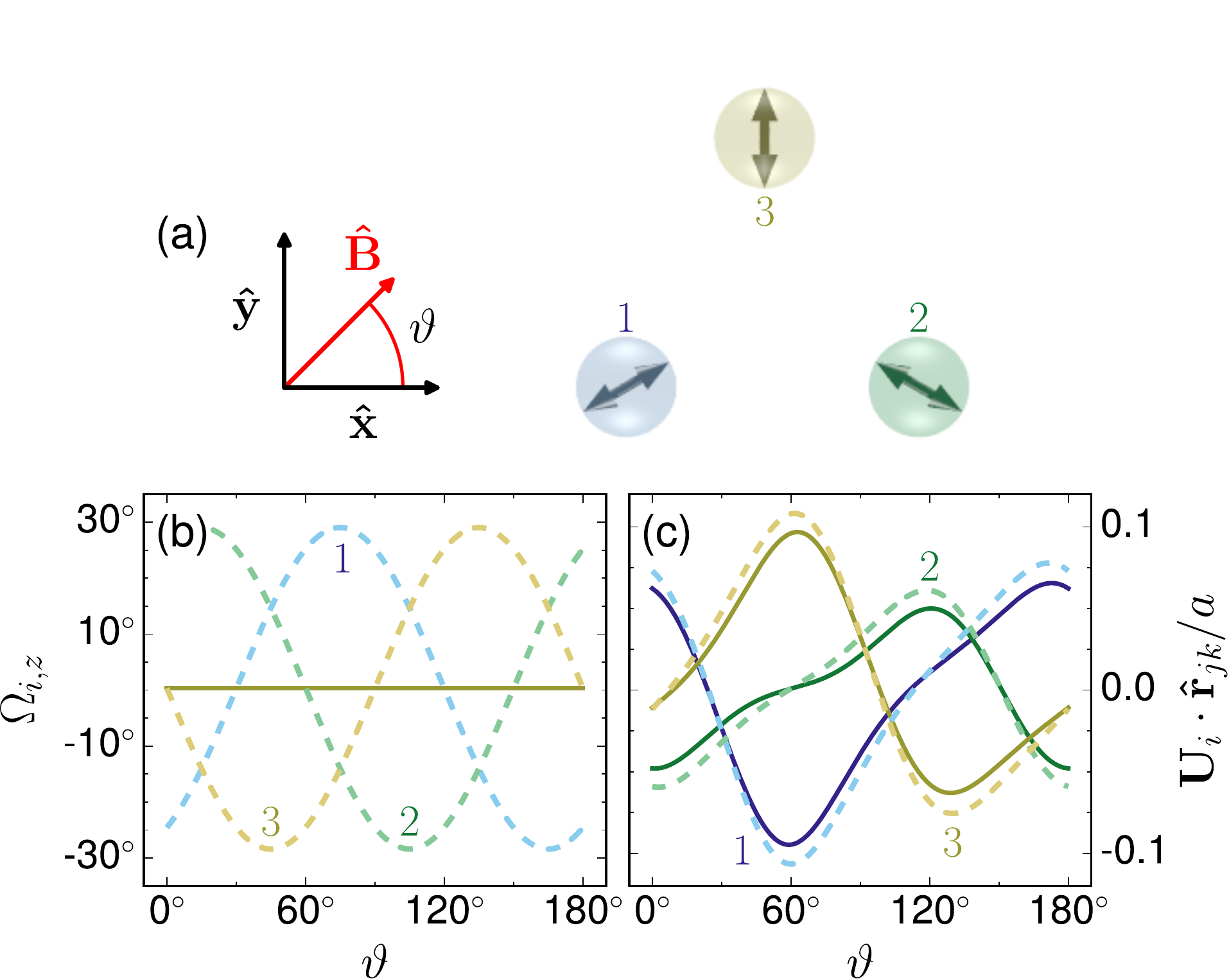}}
\caption{
Similar to Fig.~\ref{fig5} but now for a three-particle system.
(a) Schematic illustration of the initial spatial configuration of the three-particle system in Ref.~\onlinecite{puljiz2016forces} and the orientations of the added initial anisotropy axes.
Here, the anisotropy axes $\mathbf{\hat{n}}_i$ are rotated with respect to each other by $120^\circ$, with $\mathbf{\hat{n}}_3$ along $\mathbf{\hat{y}}$.
(b) Plot of the $z$-components of the rotation vectors $\boldsymbol{\Omega}_i$. Again, in this configuration all rotations take place in the $xy$-plane.
The individual curves are phase-shifted with respect to each other according to the initial shifted orientations of the anisotropy axes.
(c) Projection of the displacements $\mathbf{U}_i$ onto the interparticle unit vector $\mathbf{\hat{r}}_{jk}$ set by the respective other particles [with $(i,j,k) \in \{(1,2,3),(2,3,1),(3,1,2)\}$].
The induced torques amplify the magnitudes of displacements in the directions $\mathbf{\hat{r}}_{jk}$ (dashed lines).
Due to small deviations of the configuration from a perfect equilateral triangle \cite{puljiz2016forces}, the curves are not simply phase-shifted with respect to each other.
}
\label{fig8}
\end{figure}

\begin{figure}
\centerline{\includegraphics[width=\columnwidth]{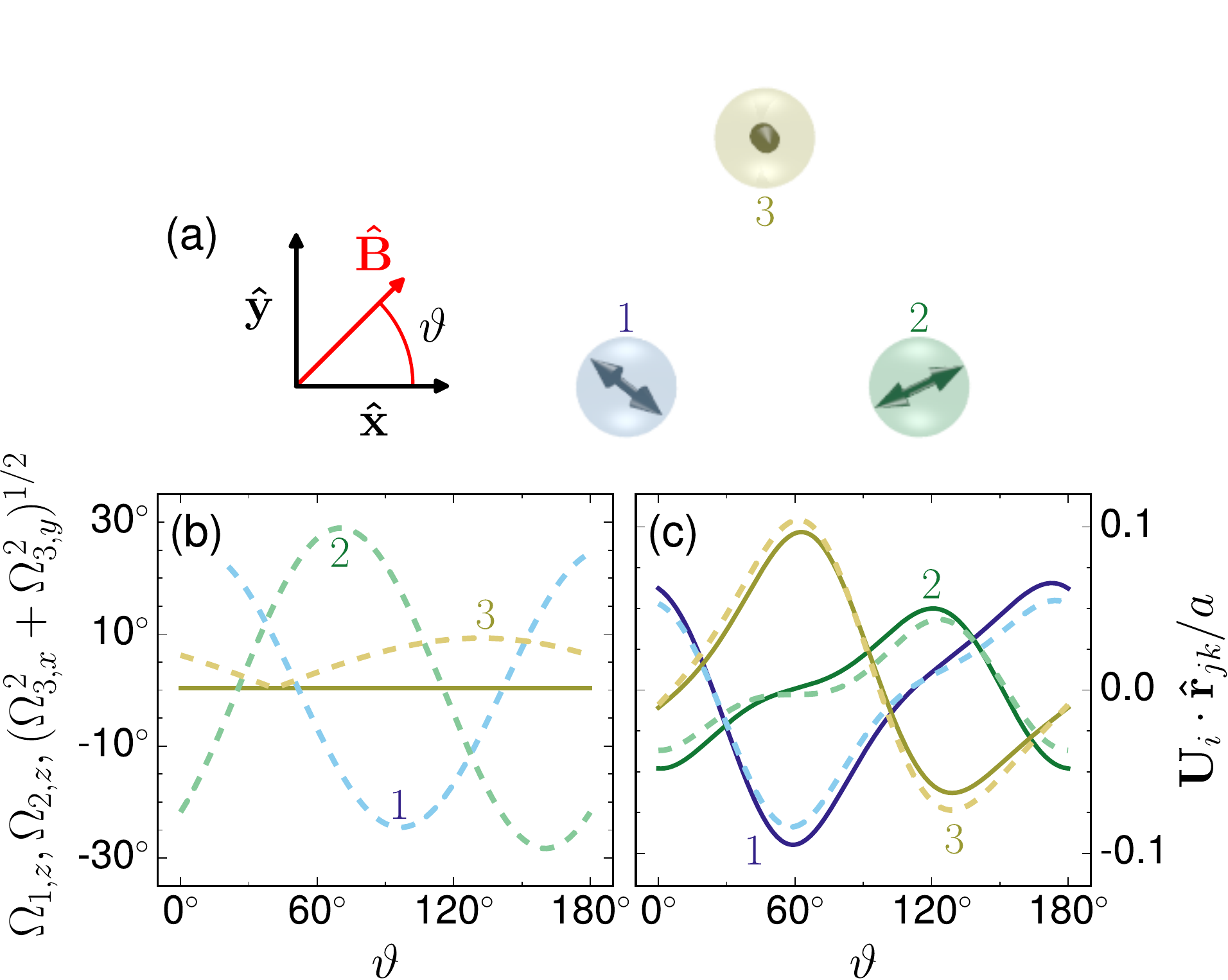}}
\caption{
The same as in Fig.~\ref{fig8}, but now the anisotropy axes are oriented ``randomly" in all three dimensions as indicated in (a). 
(b) Here, components of the rotations $\boldsymbol{\Omega}_i$ are plotted as $\Omega_{1,z}, \Omega_{2,z}, [(\Omega_{3,x})^2+(\Omega_{3,y})^2]^{1/2}$. Since initially $\mathbf{\hat{n}}_3$ is almost oriented along $\mathbf{\hat{z}}$, the torques $\mathbf{T}_3$ and therefore the rotations $\boldsymbol{\Omega}_3$ are relatively small compared to those of particles $1$ and $2$, which have a larger projection onto the $xy$-plane.
Moreover, the symmetry of Fig.~\ref{fig8} (b) does not exist anymore.
(c) Due to the additional torques, the projections $\mathbf{U}_i\cdot\mathbf{\hat{r}}_{jk}$ for particles $1$ and $2$ are reduced (dashed lines), whereas
the result for particle $3$ remains qualitatively the same as in Fig.~\ref{fig8}.
Here, additional displacements out of the $xy$-plane occur (not shown).
}
\label{fig9}
\end{figure}

Confining ourselves to the sole effect of induced forces between the embedded particles, we have in a previous work determined the resulting coupled translations \cite{puljiz2016forces}. For this purpose, we considered an example system of identical spherical paramagnetic particles that were embedded in a planar configuration into a soft elastic polymeric gel matrix. Then, an external magnetic field was applied and rotated within the configurational plane. In this way, magnetic interactions between the particles were induced and tuned by rotating the field. The elevated amplitude of the magnetic field caused a close-to-saturation magnetization of the particles. Thus, the induced magnetic dipole moments $\mathbf{m}=m\mathbf{\hat{m}}$ ($m=|\mathbf{m}|$) of the particles could be considered identical and aligned along the external magnetic field. Then, the magnetic dipole--dipole force on a particle $i$ is given by \cite{jackson1962classical}
\begin{equation}\label{magnetic_force}
	\mathbf{F}_i ={} -\frac{3\mu_0 m^2}{4\pi}\sum_{\scriptsize\begin{aligned}j\! &=\! 1 \\[-4pt] j\! &\ne \! i\end{aligned}}^N \frac{5 \mathbf{\hat{r}}_{ij} (\mathbf{\hat{m}}\cdot\mathbf{\hat{r}}_{ij})^2 - \mathbf{\hat{r}}_{ij} - 2 \mathbf{\hat{m}} (\mathbf{\hat{m}}\cdot\mathbf{\hat{r}}_{ij})}{r_{ij}^4},
\end{equation} 
where $\mu_0$ is the magnetic vacuum permeability and $N$ the total number of particles. 
We then evaluated the coupled translations resulting for the magnetized particles in response to the induced magnetic forces. Based on the magnetic nature of the particles and their size, this pure focus on induced forces and resulting translations was justified.

Here, we consider the effect of additional torques applied to the particles. The translationally and rotationally coupled situation is analyzed. We demonstrate for some minimal example configurations how the additional torques and rotational couplings modify our previous results.

For illustration, we assume the following idealized model situation. Again, we consider identical spherical magnetizable particles with no-slip surface conditions. As before, a strong external magnetic field shall be applied that saturates the magnetization of the particles and always keeps their magnetic moments oriented along the external field. However, the particles shall now be magnetically anisotropic. More precisely, we assume uniaxial magnetic anisotropy. That is, an energetic penalty arises whenever the nonpolar axis $\mathbf{\hat{n}}_i$ of magnetic anisotropy of each particle $i$ is not aligned parallel to the direction $\mathbf{\hat{m}}=\mathbf{\hat{B}}$ of the external magnetic field.
Assuming particles of this kind and following the idealized Stoner-Wohlfarth model \cite{stoner1948mechanism}, the energetic penalty for misalignment is expressed as
\begin{equation}\label{eqSW}
E_\text{SW} = K V_\text{S}\left[ 1-\left(\mathbf{\hat{n}}_i\cdot\mathbf{\hat{B}}\right)^2 \right]. 
\end{equation}
In general, $V_\text{S}$ denotes the volume of each particle and the anisotropy parameter $K$ quantifies the strength of its uniaxial magnetic anisotropy. Its magnitude may vary significantly with the magnetic nature of the particles and their shape. One factor is the type of internal lattice structure in the particles that may cause the magnetic anisotropy \cite{chen1998size,roeder2015magnetic}. 
Moreover, an elongated, e.g., rod-like shape of the particles may likewise cause magnetic uniaxiality \cite{bender2011synthesis,bender2013determination}.
Since here we are considering spherical particles, our uniaxiality must be due to a magnetocrystalline anisotropy axis.
Below, we set the rescaled relative strengths of magnetic interactions $m^2\mu_0/\mu a^6=22.5\times 10^3$ and $24.5\times 10^3$ for the considered two- and three-particle systems, respectively, corresponding to the experimental parameters in our previous study \cite{puljiz2016forces}. Moreover, we then choose a comparatively low value for the rescaled anisotropy parameter of $K/\mu=3$ \cite{skomski2003nanomagnetics}.
It leads to an effect that shows up in an illustrative way when comparing to corresponding results in the absence of imposed torques. Using Eq.~(\ref{eqSW}), we can calculate the imposed torque on each particle $i$ resulting from its orientation with respect to the external magnetic field,
\begin{equation}\label{stoner-wohlfarth_torque}
	\mathbf{T}_i ={} 2KV_\text{S}(\mathbf{\hat{n}}_i\cdot\mathbf{\hat{B}})\,\mathbf{\hat{n}}_i\times\mathbf{\hat{B}}.
\end{equation}
Since the forces $\mathbf{F}_i$ change with altering interparticle distance (during the process of particle displacement), we had implemented an iterative loop to calculate the magnetic forces in the final state \cite{puljiz2016forces}.
Now, we have extended the approach to include the torques $\mathbf{T}_i$. 
Their magnitude finally decreases with progressing rotation of the anisotropy axis towards the external magnetic field.

In Figs.~\ref{fig5}--\ref{fig9} we display our results for two- and three-particle example configurations. The initial spatial arrangements, distances, and material parameters are the same as in Ref.~\onlinecite{puljiz2016forces}.
In each of Figs.~\ref{fig5}--\ref{fig9}, a schematic sketch (a) indicates the initial orientation of the magnetic anisotropy axes. 
The external magnetic field is applied in the indicated $xy$-plane (right-handed coordinate system) and rotated in a counterclockwise way, starting from $\mathbf{\hat{B}}\cdot\mathbf{\hat{x}}=1$.
The plots (b) in each figure illustrate the resulting rotations $\boldsymbol{\Omega}_i$ as functions of the angle $\vartheta=\arccos(\mathbf{\hat{B}}\cdot\mathbf{\hat{x}})$ of the magnetic field direction. 
Moreover, the plots (c) show the displacements $\mathbf{U}_i$ in distinct directions.
Continuous lines represent the results without imposed torques, whereas dashed and/or dotted lines represent the results for the torques $\mathbf{T}_i$ included.
For distinction, the curves are labeled by the corresponding particle numbers.

Several different initial configurations of the anisotropy axes were considered in Figs.~\ref{fig5}--\ref{fig9}.
All plotted quantities were calculated via Eqs.~(\ref{displaceability_matrix}), (\ref{M_II_2})--(\ref{M_0_r}), (\ref{M_II_tr})--(\ref{M_IJ_rr}), (\ref{M_tt_3}), and (\ref{magnetic_force})--(\ref{stoner-wohlfarth_torque}).
The resulting calculated rotations and their amplifying or dampening effects on the particle displacements can be qualitatively comprehended with the help of simple geometric considerations.
For example, in Fig.~\ref{fig5} (a) the anisotropy axes of both particles $1$ and $2$ are initially oriented along the $\mathbf{\hat{x}}$-axis.
From Eq.~(\ref{stoner-wohlfarth_torque}) it then follows that the torques $\mathbf{T}_1$ and $\mathbf{T}_2$ (and thus the directly induced rotations) are maximized around $\vartheta=45^\circ$, see Fig.~\ref{fig5} (b).
Both particles are therefore rotated in counterclockwise direction, thereby creating displacement fields in the surrounding matrix (see also Fig.~\ref{fig4}).
As a result of their matrix-mediated interactions, particle $2$ is pushed into the $\mathbf{\hat{y}}$-direction due to the torque $\mathbf{T}_1$, whereas particle $1$ is pushed into the ($-\mathbf{\hat{y}}$)-direction due to $\mathbf{T}_2$, see the dashed lines in comparison to the continuous lines in Fig.~\ref{fig5} (c) around $\vartheta=45^\circ$.
Overall, this leads to an amplification of the particle displacements $|U_{i,y}|$ for all $\vartheta$.

In contrast to that, in Fig.~\ref{fig6} the anisotropy axes are initially aligned along the $\mathbf{\hat{y}}$-axis, i.e., perpendicular to the anisotropy axes in Fig.~\ref{fig5}. All other parameters remain unchanged.
As a consequence, the sense of rotation of both particles is inverted with respect to the previous configuration, see Figs.~\ref{fig5} (b) and \ref{fig6} (b). This leads to a mutual damping of the magnitudes $|U_{i,y}|$, see Fig.~\ref{fig6} (c), in opposition to the previous situation in Fig.~\ref{fig5} (c).

Another example is depicted in Fig.~\ref{fig7}, where the anisotropy axis $\mathbf{\hat{n}}_2$ of particle $2$ remains the same as in Fig.~\ref{fig5}. However, $\mathbf{\hat{n}}_1$ now points out of the $xy$-plane, along the $\mathbf{\hat{z}}$-axis. That is, $\mathbf{\hat{n}}_1$ is always oriented perpendicular to the external magnetic field $\mathbf{\hat{B}}$.
From Eq.~(\ref{stoner-wohlfarth_torque}) we find that $\mathbf{T}_1=\mathbf{0}$ for all $\vartheta$.
Thus, there is no directly induced rotation of particle $1$ that would modify the overall displacement of particle $2$.
In contrast to that, the displacement $U_{1,y}$ in Fig.~\ref{fig7} (c) remains identical to $U_{1,y}$ in Fig.~\ref{fig5} (c).

In Fig.~\ref{fig8} (a), the spatial configuration of the three-particle system studied in Ref.~\onlinecite{puljiz2016forces} is illustrated.
Additional anisotropy axes are chosen such that they are rotated by $120^\circ$ with respect to each other, all of them confined to the $xy$-plane.
This is reflected by the resulting phase-shift in the torque-induced rotations, see Fig.~\ref{fig8} (b).
The displacement $\mathbf{U}_i$ of each particle $i$ in Fig.~\ref{fig8} (c) is projected onto the interparticle unit vector $\mathbf{\hat{r}}_{jk}$ between the two other particles $j$ and $k$, i.e., $(i,j,k) \in \{(1,2,3),(2,3,1),(3,1,2)\}$.
An amplification is observed for all of these displacement components. 
This can be directly inferred from the sense of the imposed rotation of each particle, see also Fig.~\ref{fig4} and Eq.~(\ref{v_i_1_rotated_expl}).

Finally, a random initial configuration of the anisotropy axes was chosen in Fig.~\ref{fig9} (a) for the same spatial configuration as in Fig.~\ref{fig8} (a).
In view of the initial set-up, we plot in Fig.~\ref{fig9} (b) the components $\Omega_{1,z}$, $\Omega_{2,z}$, and $[(\Omega_{3,x})^2+(\Omega_{3,y})^2]^{1/2}$ of the rotation vectors.
Since $\mathbf{\hat{n}}_3$ is nearly oriented along the $\mathbf{\hat{z}}$-axis, the torque $\mathbf{T}_3$ and therefore the overall rotation $\boldsymbol{\Omega}_3$ is mostly relatively weak when compared to $\mathbf{T}_1$ and $\mathbf{T}_2$, see Fig.~\ref{fig9} (b).
The orientations of the anisotropy axes of particles $1$ and $2$ can roughly be compared with those of particles $2$ and $1$ in Fig.~\ref{fig8} (a), respectively, i.e., their roles are approximately inverted.
This leads to a mutual reduction of the depicted displacement amplitudes of particles $1$ and $2$ in Fig.~\ref{fig9} (c) when the torques are included. 
In contrast to that, the depicted displacement of particle $3$ remains qualitatively the same as in Fig.~\ref{fig9} (c).

In addition to that, we have tested how the modifications above would affect the induced changes in interparticle distances that had been plotted in Ref.~\onlinecite{puljiz2016forces}.
However, the relative deviations from the situations without torques were only of the order $\sim10^{-2}$.

\section{Conclusions and outlook}\label{Section_11_conclusions}

In summary, we have presented the derivation of explicit analytical expressions to calculate from given forces and torques acting on rigid spherical inclusions in an elastic matrix their resulting coupled displacements and rotations. The surrounding elastic matrix is assumed to be an infinitely extended, homogeneous, isotropic elastic medium with stick boundary conditions on the inclusion surfaces.
Matrix deformations are induced by the forces and torques acting on the inclusions.
These deformations lead to mutual, long-ranged, matrix-mediated interactions between the rigid inclusions.
The role of such matrix-mediated interactions is implicitly contained in our resulting analytical expressions.  
Technically, to perform the derivation, the well-known approach in terms of Fax\'{e}n's theorems and the method of reflections is adapted from the field of low-Reynolds-number hydrodynamics \cite{dhont1996introduction}. 
Throughout, we have included the case of \textit{compressible} elastic environments. We summarize our results in terms of \textit{displaceability and rotateability} matrices that are functions of the given inclusion configuration only. These matrices express how given \textit{forces and torques} on the inclusions lead to their coupled \textit{displacements and rotations}.
In the considered static, linearly elastic case of non-touching inclusions, these expressions replace the need for finite-element simulations that explicitly calculate the matrix deformations between the inclusions.

As a next step, more complex inclusion geometries can be addressed. Of particular interest are elongated particles that can more directly be exposed to external torques and are also used for microrheological purposes \cite{bender2011synthesis,roeder2012shear,bender2013determination}. 
Theoretically, it should be possible to derive expressions for ellipsoidal inclusions \cite{karrila1991microhydrodynamics,kim1995faxen}, but due to the significantly more complicated structure of such expressions they may already be of limited use for practical applications. 
Long thin rods could be approximated by long chains of spheres \cite{dhont1996introduction}. Recent experiments observed a buckling of chains of spherical magnetic particles in soft gel matrices under perpendicular magnetic fields \cite{huang2016buckling}. Possibly, such behavior could likewise be interpreted more quantitatively in terms of our formalism. 
As in low-Reynolds-number hydrodynamics, more complex inclusion objects should become accessible by the raspberry model, i.e., collections of rigidly connected identical spheres that as an entity represent more complex objects \cite{lobaskin2004new,lobaskin2004electrophoretic,passow2015depolarized}.
Moreover, similarly to low-Reynolds-number hydrodynamics, the effect of system boundaries should be analyzed \cite{squires2000like,kim2006electro}.
Possibly, also hydrodynamic methods to describe more concentrated colloidal suspensions \cite{tanaka2000simulation,morris2009review} could be transferred to the case of elastic environments.

Our results will be helpful in the quantitative interpretation of microrheological experiments \cite{ziemann1994local,bausch1999measurement,waigh2005microrheology,wilhelm2008out,mizuno2007nonequilibrium,bender2011synthesis,roeder2012shear,bender2013determination}, as already indicated in our previous work \cite{puljiz2016forces}. 
In principle, they should apply to different sorts of elastic matrix environments, as long as the material appears sufficiently homogeneous and isotropic down to the scale of the probe particle. 
For example, a related picture applies to the modeling of active forces generated by and within biological cells, where particularly the effect of active force dipoles is investigated \cite{yuval2013dynamics,cohen2016elastic}.
Another field of application is to further characterize the tunability of composite materials by externally imposed fields \cite{jolly1996magnetoviscoelastic,an2003actuating,gao2004electrorheological,minagawa2005electro,filipcsei2007magnetic,ivaneyko2012effects,menzel2015tuned,odenbach2016microstructure}. For example, the change in the linear elastic moduli of magnetorheological elastomers when applying an external magnetic field could be addressed using our formalism. The method could be combined with statistical descriptions that use a probability distribution to characterize the arrangement of the inclusions in an elastic matrix \mpu{\cite{gundermann2017statistical}}. 

One strength is that larger numbers of inclusions can be handled than with simulation methods that explicitly resolve the matrix environment \cite{han2013field,spieler2013xfem,cremer2015tailoring,attaran2016modeling,metsch2016numerical,cremer2016superelastic}, at least to the accuracy given by the expansion in the particle distance \mpu{and as long as linear elasticity theory is sufficient to describe the resulting matrix deformations to the desired degree of accuracy. 
Naturally, concerning the latter point, nonlinear elastic effects arising in real materials with increasing amplitude of deformation will first quantitatively affect the results and may, for large degrees of deformation, even lead to qualitative differences in the behavior. 
Extending such formalisms as the present one to the nonlinear regime is a nontrivial future task and incomparably more involved. 
Nevertheless, as we have demonstrated, in many cases numerical and experimental results are still well reproduced. 
Thus, considering the explicit form of our resulting anlytical expressions and their efficient numerical evaluation, our approach will still be beneficial for analyzing the behavior of real materials. For example, it allows to quickly qualitatively scan the response of a multitude of different possible particle distributions and internal structural realizations in elastic composites.}
In this way, \mpu{our approach} shall help to quantitatively support the development of tunable composite materials designed for a specific requested purpose.

\acknowledgments
The authors thank Shilin Huang and G\"unter K.\ Auernhammer for stimulating discussions as well as the Deutsche Forschungsgemeinschaft for support of this work through the priority program SPP 1681, grant no.\ ME 3571/3. 

\appendix
\section*{Appendix A}\label{Appendix_fourier_greens_function}
\renewcommand{\theequation}{A.\arabic{equation}}
\setcounter{equation}{0}

Eq.~(\ref{ggb}) can be solved by Fourier forth and back transformation. The former replaces the nabla operator $\nabla$ by $i\mathbf{k}$ and the Dirac delta function $\delta(\mathbf{r}-\mathbf{r}_0)$ by $1$ in Eq.~(\ref{ggb}), 
\begin{equation}
	\lambda_{kpim}\hat{k}_m \hat{k}_p k^2\tilde{G}_{ij}(\mathbf{k})={}\delta_{jk},
\end{equation}
with the unit vector $\mathbf{\hat{k}}=\mathbf{k}/k$ in $\mathbf{k}$-space. Inserting
\begin{equation}
        \lambda_{kpim}\hat{k}_m\hat{k}_p = \mu\Bigg[\delta_{ik}+\frac{\lambda+\mu}{\mu}\hat{k}_i\hat{k}_k\Bigg]
\end{equation}
via Eq.~(\ref{clandau}), we can solve for the Green's function in Fourier space:
\begin{eqnarray}
	\mathbf{\hspace{.02cm}\underline{\hspace{-.02cm}\tilde{G}}}(\mathbf{k}) 
	&={} &\frac{1}{\mu k^2}\Bigg[\mathbf{\underline{\hat{I}}}-\frac{\lambda+\mu}{\lambda+2\mu} \mathbf{\hat{k}}\mathbf{\hat{k}} \Bigg]\notag\\
	&={} &\frac{1}{\mu k^2}\Bigg[\mathbf{\underline{\hat{I}}}-\frac{1}{2(1-\nu)} \mathbf{\hat{k}}\mathbf{\hat{k}} \Bigg], \label{green_fourier}
\end{eqnarray}
with $\mathbf{\underline{\hat{I}}}$ the identity matrix and $\mathbf{\hat{k}}\mathbf{\hat{k}}$ a dyadic product.
Next, we transform back to real space,
\begin{widetext}
\begin{eqnarray}
	\mathbf{\hspace{.02cm}\underline{\hspace{-.02cm}G}}(\mathbf{r}) &={} &\frac{1}{(2\pi)^3} \int_{0}^{2\pi}\mathrm{d}\varphi\int_{0}^{\pi}\mathrm{d}\vartheta\sin\vartheta\int_{0}^{\infty}\mathrm{d}k k^2 e^{i\mathbf{k}\cdot\mathbf{r}} \mathbf{\hspace{.02cm}\underline{\hspace{-.02cm}\tilde{G}}}(\mathbf{k}) \notag\\
	&={} & \frac{1}{(2\pi)^3\mu} \int_{0}^{2\pi}\mathrm{d}\varphi\int_{0}^{\pi}\mathrm{d}\vartheta\sin\vartheta \int_{0}^{\infty}\mathrm{d}k\, e^{ikr\cos\vartheta} \Bigg[\mathbf{\underline{\hat{I}}}-\frac{1}{2(1-\nu)} \mathbf{\hat{k}}\mathbf{\hat{k}}\Bigg].\label{green_zwischenschritt}
\end{eqnarray} 
The Dirac delta function is linked to its Fourier transform via 
\begin{equation}
	\int_{-\infty}^{\infty}\mathrm{d}k\, e^{ikx} ={} \int_{-\infty}^{\infty}\mathrm{d}k \Big[ \cos(kx) + i \sin(kx)\Big] ={} 2\pi\delta(x).\label{dirac_fourier}
\end{equation}
Keeping this in mind, the $k$-integral in the second line of Eq.~(\ref{green_zwischenschritt}) is reformulated: 
\begin{eqnarray}
	\int_{0}^{\infty}\mathrm{d}k\, e^{ikr\cos\vartheta} &={} &\int_{0}^{\infty}\mathrm{d}k \cos(kr\cos\vartheta) + \int_{0}^{\infty}\mathrm{d}k \,i \sin(kr\cos\vartheta) \notag\\
	&={} & \frac{1}{2}\int_{-\infty}^{\infty}\mathrm{d}k \cos(kr\cos\vartheta) + \int_{0}^{\infty}\mathrm{d}k \,i \sin(kr\cos\vartheta) \notag\\
	&={} & \frac{1}{2}\int_{-\infty}^{\infty}\mathrm{d}k \Big[\cos(kr\cos\vartheta) +  i \sin(kr\cos\vartheta) \Big] - \frac{1}{2}\int_{-\infty}^{0} \mathrm{d}k\, i \sin(kr\cos\vartheta)\notag\\
	&={} & \pi\delta(r\cos\vartheta) - \frac{1}{2}\int_{-\infty}^{0} \mathrm{d}k\, i \sin(kr\cos\vartheta).
\end{eqnarray}
We find that the second term in the last line of the previous expression does not contribute. Upon inserting it into Eq.~(\ref{green_zwischenschritt}), it leads to
\begin{equation}
	\int_{0}^{2\pi}\mathrm{d}\varphi\int_{0}^{\pi}\mathrm{d}\vartheta\sin\vartheta \int_{-\infty}^{0} \mathrm{d}k \sin(kr\cos\vartheta) \Bigg[\mathbf{\hat{I}}-\frac{1}{2(1-\nu)} \mathbf{\hat{k}}\mathbf{\hat{k}}\Bigg].
\end{equation}\end{widetext}
Substituting $u=\cos\vartheta$ and $-\mathrm{d}u ={} \sin\vartheta\,\mathrm{d}\vartheta$, it can easily be seen that the first term in the square brackets leads to an odd function of $u$ and therefore vanishes upon integration over $\mathrm{d}u$ from $u=1$ to $-1$.
Calculating for the second term in the square brackets all matrix components $\hat{k}_i \hat{k}_j$ explicitly by inserting the components of $\mathbf{\hat{k}}$, the second term is found to vanish as well.

Thus, for the remainder of Eq.~(\ref{green_zwischenschritt}), we obtain
\begin{eqnarray}
	\mathbf{\hspace{.02cm}\underline{\hspace{-.02cm}G}}(\mathbf{r}) &={} & \frac{1}{8\pi^2\mu r} \int_{0}^{2\pi}\mathrm{d}\varphi\int_{-1}^{1}\mathrm{d}u\, \delta(u) \Bigg[\mathbf{\underline{\hat{I}}}-\frac{1}{2(1-\nu)} \mathbf{\hat{k}}\mathbf{\hat{k}}\Bigg]\notag\\
	&={} &\frac{1}{8\pi^2\mu r} \int_{0}^{2\pi}\mathrm{d}\varphi \Bigg[\mathbf{\underline{\hat{I}}}-\frac{1}{2(1-\nu)} \mathbf{\hat{k}}\mathbf{\hat{k}}\Bigg]\Bigg|_{\mathbf{\hat{k}}\cdot\mathbf{r}=0},\label{int_G_F}
\end{eqnarray}
with the condition $\mathbf{\hat{k}}\perp \mathbf{r}$ arising from the delta function. Thus, $\mathbf{\hat{k}}$ can be expressed as
\begin{equation}\label{k_alpha_beta}
	\mathbf{\hat{k}} ={} \boldsymbol{\hat{\upalpha}}\cos\varphi+\boldsymbol{\hat{\upbeta}}\sin\varphi,
\end{equation}
with the constant unit vectors $\boldsymbol{\hat{\upalpha}}$ and $\boldsymbol{\hat{\upbeta}}$, $\boldsymbol{\hat{\upalpha}}\perp\boldsymbol{\hat{\upbeta}}$, and $\boldsymbol{\hat{\upalpha}}\perp\mathbf{r}\perp\boldsymbol{\hat{\upbeta}}$. 
Then, $\boldsymbol{\hat{\upalpha}}$, $\boldsymbol{\hat{\upbeta}}$, and $\mathbf{\hat{r}}=\mathbf{r}/r$ form an orthonormal basis and we can write
\begin{equation}
	\boldsymbol{\hat{\upalpha}}\boldsymbol{\hat{\upalpha}}+\boldsymbol{\hat{\upbeta}}\boldsymbol{\hat{\upbeta}}+\mathbf{\hat{r}}\mathbf{\hat{r}} ={} \mathbf{\underline{\hat{I}}}.
\end{equation}
Inserting Eq.~(\ref{k_alpha_beta}) into Eq.~(\ref{int_G_F}), we evaluate the remaining integral over $\mathrm{d}\varphi$ and obtain
\begin{eqnarray}
	\mathbf{\hspace{.02cm}\underline{\hspace{-.02cm}G}}(\mathbf{r}) &={} &\frac{1}{8\pi\mu r}\Bigg[ 2\mathbf{\underline{\hat{I}}} - \frac{1}{2(1-\nu)}\big( \boldsymbol{\hat{\upalpha}}\boldsymbol{\hat{\upalpha}} + \boldsymbol{\hat{\upbeta}}\boldsymbol{\hat{\upbeta}} \big) \Bigg]\notag\\
&={}&	\frac{1}{8\pi\mu r}\Bigg[ 2\mathbf{\underline{\hat{I}}} - \frac{1}{2(1-\nu)}\big( \mathbf{\underline{\hat{I}}}- \mathbf{\hat{r}}\mathbf{\hat{r}} \big) \Bigg].
\end{eqnarray}
Finally, combining \mpu{the prefactors of $\mathbf{\underline{\hat{I}}}$} leads to the expression for the elastic Green's function in Eq.~(\ref{greens_function}).

\section*{Appendix B}\label{Appendix_principal_value}
\renewcommand{\theequation}{B.\arabic{equation}}
\setcounter{equation}{0}
Our goal is to evaluate the integral
\begin{equation}\label{appendix_1}
	\frac{1}{2}\int\limits_{-\infty}^{\infty}\mathrm{d}k\,\frac{\sin(ka)}{ka}e^{ik\mathbf{\hat{k}}\cdot\mathbf{r}}
\end{equation}
appearing in Eq.~(\ref{SIN_KA}).
For this purpose, we rewrite the expression by substituting $z=ka$:
\begin{widetext}
\begin{equation}
	\frac{1}{2}\int\limits_{-\infty}^{\infty}\mathrm{d}k\,\frac{\sin(ka)}{ka}e^{ik\mathbf{\hat{k}}\cdot\mathbf{r}}
	={} \frac{1}{4ia} \int\limits_{-\infty}^{\infty}\mathrm{d}z\,\frac{1}{z}\left[e^{iz\left( 1+\frac{\mathbf{\hat{k}}\cdot\mathbf{r}}{a} \right)} - e^{iz\left( -1+\frac{\mathbf{\hat{k}}\cdot\mathbf{r}}{a} \right)} \right].\label{app_all_1}
\end{equation}
\end{widetext}
The evaluation can be accomplished in a straightforward way by using contour integration in the complex $z$-plane.
We start by considering only the first term on the right-hand side and define the function
\begin{equation}
	f(z)={} \frac{1}{z}e^{iz\left(1+\frac{\mathbf{\hat{k}}\cdot\mathbf{r}}{a}\right)}.
\end{equation}
Depending on the value of $\mathbf{\hat{k}}\cdot\mathbf{r}/a$, the integration path is amended on a case-by-case basis over a semicircle of infinite radius $R$ in either the upper or the lower complex $z$-half-plane.
Starting with $\mathbf{\hat{k}}\cdot\mathbf{r}/a>-1$, the integration path is closed in the upper $z$-half-plane.
According to Cauchy's integral theorem, in our case all closed integration paths that do not contain the origin are zero, therefore
\begin{widetext}
\begin{equation}
	0={} \oint \mathrm{d}z\,f(z) ={} \lim_{R\rightarrow\infty}\left[\,\,\int\limits_{-R}^{-\varepsilon}\mathrm{d}z\,f(z) - \int\limits_{\mathcal{C}_\varepsilon}\mathrm{d}z\, f(z) +\int\limits_{\varepsilon}^{R}\mathrm{d}z\,f(z) + \int\limits_{\mathcal{C}_R}\mathrm{d}z\,f(z) \right],
\end{equation}
\end{widetext}
with
	$\mathcal{C}_\varepsilon=\{\varepsilon e^{i\varphi}\,|\,0\le\varphi\le\pi\}$ and  $\mathcal{C}_R=\{R e^{i\vartheta}\,|\,0\le\vartheta\le\pi\}$.
The integral over the path $\mathcal{C}_R$ vanishes for $R\rightarrow\infty$. Combining these relations with the principal value,
\begin{equation}
	\mathcal{P}\int (\dots\!\,)={} \lim_{\varepsilon\searrow 0}\left[\,\,\int\limits_{-\infty}^{-\varepsilon}(\dots\!\,) + \int\limits_{\varepsilon}^{\infty}(\dots\!\,)\right],
\end{equation}
we obtain in this first case
\begin{equation}
	\mathcal{P}\int\limits_{-\infty}^\infty\mathrm{d}z\, f(z) ={} \lim\limits_{\varepsilon\searrow 0}i\int\limits_{0}^{\pi}\mathrm{d}\varphi\,e^{i\varepsilon e^{i\varphi}\left(1+\frac{\mathbf{\hat{k}}\cdot\mathbf{r}}{a}\right)} ={} i\pi.\label{PRINCIPAL_I}
\end{equation}

\mpu{Similarly,} for $\mathbf{\hat{k}}\cdot\mathbf{r}/a<-1$ we amend the integration path over the semicircle of infinite radius in the lower $z$-half-plane and obtain for the principal value
\begin{equation}
	\mathcal{P}\int\limits_{-\infty}^{\infty}\mathrm{d}z\,f(z) ={} 
	-\lim_{\varepsilon\searrow 0}i\int\limits_{\pi}^{2\pi}\mathrm{d}\varphi\, e^{i\varepsilon e^{i\varphi}\left(1+\frac{\mathbf{\hat{k}}\cdot\mathbf{r}}{a}\right)} ={} -i\pi.\label{PRINCIPAL_II}
\end{equation}

An analogous procedure for the second term on the right-hand side of Eq.~(\ref{app_all_1}) yields
\begin{equation}
	\mathcal{P}\int\limits_{-\infty}^{\infty}\mathrm{d}z\,\frac{1}{z}e^{iz\left( -1+\frac{\mathbf{\hat{k}}\cdot\mathbf{r}}{a} \right)} ={} \begin{cases}
		i\pi,\qquad &\text{for }\frac{\mathbf{\hat{k}}\cdot\mathbf{r}}{a} > 1,\\[0.5em]
		-i\pi,\qquad &\text{for }\frac{\mathbf{\hat{k}}\cdot\mathbf{r}}{a} < 1.
	\end{cases}\label{PRINCIPAL_III}
\end{equation}

Inserting Eqs.~(\ref{PRINCIPAL_I})--(\ref{PRINCIPAL_III}) into Eq.~(\ref{app_all_1}) finally leads to \cite{dhont1996introduction}
\begin{equation}
	\frac{1}{2}\int\limits_{-\infty}^{\infty}\mathrm{d}k\,\frac{\sin(ka)}{ka}e^{ik\mathbf{\hat{k}}\cdot\mathbf{r}} ={} \begin{cases}
	\frac{\pi}{2a}, &\text{for }-1<\frac{\mathbf{\hat{k}}\cdot\mathbf{r}}{a}<1,\\
	0, &\text{otherwise}.
	\end{cases}
\end{equation}
\\

\end{document}